\documentclass{article} 
\usepackage{iclr2027_conference,times}

\usepackage{amsmath,amsfonts,bm}

\def\eqref#1{equation~\ref{#1}}

\def\1{\bm{1}}

\DeclareMathAlphabet{\mathsfit}{\encodingdefault}{\sfdefault}{m}{sl}
\SetMathAlphabet{\mathsfit}{bold}{\encodingdefault}{\sfdefault}{bx}{n}

\usepackage{hyperref}
\usepackage{url}
\usepackage[detect-all]{siunitx}
\DeclareSIUnit{\calorie}{cal}
\DeclareSIUnit\angstrom{\protect \text {Å}}

\usepackage{cleveref}
\usepackage{xcolor}
\usepackage{booktabs}
\usepackage{graphicx}
\usepackage{colortbl} 
\usepackage{longtable, booktabs}
\usepackage[section]{placeins}

\definecolor{replonly}{HTML}{CDEEFB}

\title{mdRNAcat: A Molecular Dynamics RNA \newline{}Catalog of REST2 Conformational Ensembles}

\author{
\shortstack[l]{
    Anton E. Dorn$^{1,2}$,
    Emile de Bruyn$^{1}$,
    Fabrice von der Lehr$^{3}$ \\
    Philipp Knechtges$^{3}$,
    Stefan Kesselheim$^{1,4}$,
    Alexander Schug$^{5}$
} \\[0.8em]
$^{1}$ Jülich Supercomputing Centre, Forschungszentrum Jülich (FZJ), \\
$^{2}$ Faculty of Physics, Georg August University of Göttingen, \\
$^{3}$ Institute of Software Technology, Germany Aerospace Center (DLR), \\ 
$^{4}$ Department of Physics, University of Cologne \\
$^{5}$ Scientific Computing Center, Karlsruhe Institute of Technology (KIT) \\[0.5em]
\texttt{\{a.dorn,e.de.bruyn,s.kesselheim\}@fz-juelich.de} \\[0.2em]
\texttt{\{fabrice.lehr,philipp.knechtges\}@dlr.de} \\[0.2em]
\texttt{alexander.schug@kit.edu}
}

\iclrfinalcopy 

\begin{document}

\maketitle
\lhead{Preprint} 

\begin{abstract}
Learning molecular models of RNA requires simulation data that capture conformational variability and provide consistent physical training signals. Here, we introduce mdRNAcat, a curated dataset of all-atom molecular dynamics (MD) simulations of 48 RNA systems selected to promote sequence diversity. Each system is simulated using replica-exchange with solute scaling (REST2), with 32 Hamiltonian states and production lengths of 200–300 ns per state. The dataset provides atomic coordinates, instantaneous forces, and forces reevaluated under the unscaled reference Hamiltonian, together with structural annotations and replica-exchange metadata. Coordinates and forces are available at 1 ps intervals for the reference state and 100 ps intervals across all states in a standardized HDF5 format. We characterize conformational sampling, exchange behavior, base pairing, and sugar-pucker distributions to document the properties of the simulated ensembles. mdRNAcat provides a reusable resource for developing and evaluating data-driven RNA models, including coarse-grained force fields.
\end{abstract}

\section{Introduction} \label{sec:introduction}

At the sub-cellular scale, life is an intricate choreography of biomolecules in motion, whose understanding requires to know both their structure and dynamics.
Despite major advances in experimental techniques, many important biomolecules remain structurally unresolved, with their conformational dynamics, at best, partially characterized.
In principle, molecular dynamics (MD) simulations provide atomistic access to such dynamics, but the relevant time scales are often out of reach due to prohibitive computational expenses (\cite{hollingsworth2018molecular}).
Recently, machine learning methods have made substantial progress in structure prediction, such as Alphafold 3 (\cite{alphafold3}) but also for learning molecular interactions and dynamics (\cite{kovacs2025mace}). However, much of this progress and the underlying training data have centered on proteins rather than RNA.
This is in part due to differences in available data. Machine learning methods rely on available data to achieve the aforementioned results. For proteins, large collections of experimentally determined structures (\cite{berman2000protein}), MD data sets, such as mdCATH and ATLAS (\cite{mirarchi2024mdcath,vander2024atlas}), and data sets of quantum-mechanical properties (\cite{levine2025open,eastman2024nutmeg}) are available.
In contrast, RNA lacks a comparably large and systematically generated MD resource. While curated RNA structure collections and nucleic-acid simulation repositories exist \cite{nucleoseeker,hospital2016bignasim}, they do not provide the combination of scale, homogeneous protocols, large-scale conformational sampling, and consistently available force information targeted by mdRNAcat. To our knowledge, no existing RNA MD dataset combines these properties. This gap is particularly relevant given the growing interest in data-driven RNA modeling \cite{upadhyay2026sequence} and the need for large, consistent training datasets for coarse-grained machine-learned force fields \cite{RNA_CG_FF_review}.

More precisely, while existing molecular simulation repositories are valuable resources, their use for systematic data-driven analyses is limited when trajectories originate from heterogeneous simulation protocols \cite{vander2024atlas,kouba2023machine}. Variations in force fields, solvent models, simulation conditions, sampling strategies, and trajectory lengths can introduce systematic differences that are difficult to separate from genuine molecular variation. Moreover, available observables are often inconsistent across repositories and datasets; in particular, atomic forces, which are essential for learning molecular interactions and fitting force fields, are not routinely retained \cite{mirarchi2024mdcath}. For studying RNA conformational ensembles and developing transferable data-driven models, an ideal dataset should therefore combine broad coverage of relevant RNA systems with a homogeneous simulation protocol, enhanced conformational sampling\cite{rest2,bussi2020metadynamics}, and a consistent set of physical observables, including both coordinates and forces.

\begin{figure}[h]
\begin{center}
\includegraphics[width=0.8\textwidth]{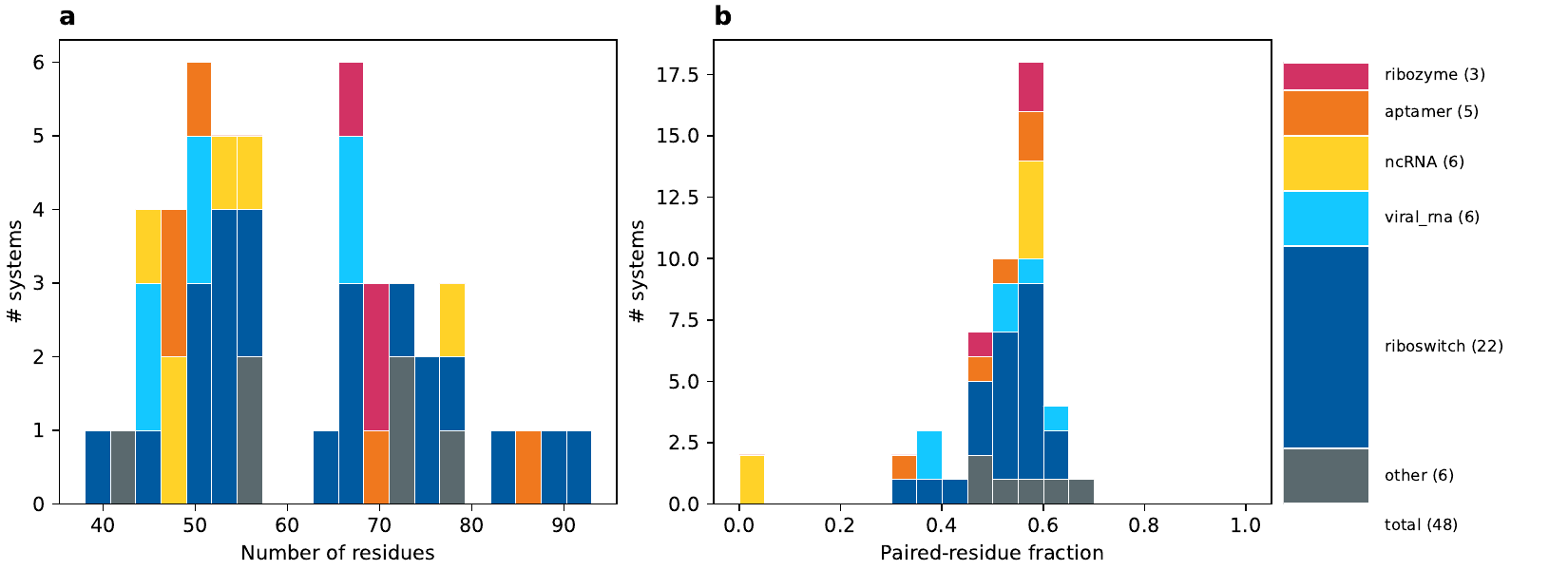}
\end{center}
\caption{(\textbf{a}) Distribution of the number of residues per RNA system in the dataset. (\textbf{b}) Distribution of the percentage of paired residues in the dataset; most systems display similar behavior, apart from one RNA system in an extended state. Colors indicate different RNA domains.}
\label{plot:histograms}
\end{figure}

To address these limitations, we present mdRNAcat, a homogeneous, enhanced-sampling MD dataset designed for systematic analysis and data-driven modeling of RNA conformational ensembles (Figure~\ref{plot:histograms} and Table~\ref{table:dataset_stats}). The dataset is intended to support tasks such as force matching, learned molecular potentials, coarse-grained model development, and representation learning of RNA conformational ensembles.

Our main contributions are:
\begin{itemize}
    \item We \textbf{generate a systematically curated set} of 48 sequence-, structurally and functionally diverse RNA systems simulated under a single, consistent all-atom MD protocol with REST2 enhanced sampling, \textbf{spanning diverse regions of RNA sequence space while broadening conformational sampling beyond conventional MD}.

    \item We provide a \textbf{rich and standardized set of physical training signals}, including atomic coordinates, instantaneous forces, and reference-Hamiltonian forces for configurations sampled across REST2 states, together with structural annotations and replica-exchange metadata \textbf{in a common HDF5 representation}.

    \item We \textbf{quantitatively characterize and validate mdRNAcat} in terms of sequence-space coverage, enhanced conformational sampling, replica-exchange statistics, and structural properties, \textbf{providing dataset-wide quality-control measures for downstream data-driven RNA modeling}.
\end{itemize}


\begin{table}[t]
\caption{Statistical overview of the dataset.}
\label{table:dataset_stats}
\centering
\small
\sisetup{
    group-separator = {,},
    group-minimum-digits = 4
}
\begin{tabular}{
    @{}
    l
    S[table-format=4.1]
    @{\;}l
    @{}
}
\toprule
Total RNA systems                                     & 48    & \\
Systems with \qty{200}{\nano\second} simulation time  & 45    & \\
Systems with \qty{300}{\nano\second} simulation time  & 3     & \\
Replicas                                              & 32    & \\
State trajectories                                    & 1536  & \\
Total sampled time                                    & 316.8 & \unit{\micro\second} \\
Total sampled time at \qty{300}{\kelvin}              & 9.9   & \unit{\micro\second} \\
Average atoms per system                              & 1955  & \\
Average residues per system                           & 61    & \\
Total file size                                       & 699   & \unit{\giga\byte} \\
\bottomrule
\end{tabular}
\end{table}

\section{Dataset Requirements} \label{sec:dataset_req}

To achieve the goals for the dataset, we set a number of requirements that the data should adhere to.

\begin{itemize}
    \item \textit{Reproducibility:} The \texttt{.pdb} files and GROMACS topology files are included in the dataset; the processing and simulation protocols are described in detail in the Appendix \ref{appendix:simulation_params}. This ensures comparable and consistent future work on this dataset.
    \item \textit{Interoperability:} The dataset is saved in the HDF5 format as a widely used and supported data format. The formatting is kept as similar to mdCATH (\cite{mirarchi2024mdcath}) as possible, to ensure reusability of code.
    \item \textit{Accessibility:} The files are annotated with all relevant meta data on the trajectories as well as derived parameters such as radius of gyration or sugar pucker angle. The trajectories contain coordinates as well as the forces and for higher temperature replicas also the recalculated forces at the lowest temperature.
    \item \textit{High quality data:} For accurate learning of the RNA conformation, an accurate representation in the training set is vital. Thus we used state-of-art force-field, sampling methods and simulation code to generate mdRNAcat. Critically, the simulations should be homogeneously run.
    \item \textit{Sequence space coverage:} Sequence-space coverage was quantified against a filtered set of all available RNA sequences with structures in the PDB, using the analysis described in Section \ref{subsec:seq-val}.
    \item \textit{Conformation space coverage:} To also sample the conformation space widely, we used state-of-art a replica exchange algorithm in REST2 to find and sample conformations further from the global minimum.
\end{itemize}

\section{Methods} \label{sec:methods}

\subsection{Sequence set generation}
The data set is based on experimentally resolved sequences from the RCSB Protein Data Bank  (PDB) (\cite{berman2000protein}). With NucleoSeeker (\cite{nucleoseeker}), a set of sequences was searched that have a pairwise sequence identity of less than 50\,\% and contain the highest number of sequences in the set, thus covering as much of the known RNA structure space as possible. Further criteria were structures containing only RNA and a sequence length of at least 40 residues to ensure that short common sequences do not disallow most of the sequence space due to high sequence identity. Additionally, only X-Ray structures with a resolution below \qty{3.6}{\angstrom} were initially considered. This led to a set containing 105 RNA sequences and structures with length between 40 and 192 residues. From this set 43 structures were chosen and partly edited to be readable for GROMACS with truncation of modified regions or insertion of missing atoms. The remaining 62 structures were not usable for GROMACS, out of scope for the chosen simulation setup and available compute resources,\footnote{Scope cutoff is not solely based on residue number of the system, the system sizes were tested in short simulations to produce good replica exchange statistics, see Section \ref{subsec:repex-val}} or de-prioritized in favor of other systems. Structures were not usable for GROMACS because of non-recoverable modifications that were not implemented in the used forcefield or large gaps in the structure that were too large to reconstruct. Lastly, 5 NMR-structures were manually added to broaden the sequence space, yielding a total of 48 structures. A detailed list with all systems considered and used can be found in the Appendix \ref{appendix:pdb_list}. 

\subsection{Simulation} \label{subsec:simulation}

The trajectories contained in mdRNAcat were generated with all-atom Molecular Dynamics (MD) simulation. MD simulation are physics-based approach and common tool for computational analysis of biomolecular system (\cite{hollingsworth2018molecular}). For MD simulations, the system needs to be described with physical parameters such as atom charge and bond strengths, which are called force fields (FF). These are used to calculate frame-wise the forces and thus velocities and positions for each particle in the simulated system. Because the energy barriers between different minima in the conformation space of biomolecule are commonly large, often enhanced sampling is used to improve MD simulations (\cite{abrams2013enhanced}). In this work, we used replica exchange with solute scaling (REST2) algorithm (\cite{rest2}) implemented via PLUMED (\cite{PLUMED1,PLUMED2,PLUMED3}). REST2 does not scale the temperature of the replica systems, but modifies their Hamiltonian and scales the solute-solute and solute-solution interactions. As a consequence, only the lowest temperature replica is physically correctly simulating its temperature, but since it only scales the interaction, it also allows for recalculating the forces correctly for all replicas as if they were at the lowest temperature.

In this work, GROMACS (\cite{abraham2015gromacs}) was used for all pre-processing and simulations of the systems. For the RNA parameters the tAMBER99SB-ildn-star force field with parmbsc0 corrections (\cite{case2025recent,perez2007refinement,guy2012single}) with reparameterization  from \cite{tan2018rna} and  phosphate parameters from \cite{steinbrecher2012revised} was used. Solvation was done in cubic boxes with TIP4PD water(\cite{piana2015water}) and a padding of 10 to \qty{20}{\angstrom} depending on the system. The systems were neutralized and ionized with K$^+$ and Cl$^-$ ions at 0.15\,M concentration with CHARMM22-ions (\cite{mackerell1998all}). A number of cations equal to half the number of residues in the system was manually and randomly placed closer to the negatively charged solute to achieve faster equilibration. Additionally particle-mesh Ewald (PME) with a cutoff of \qty{12}{\angstrom} for long-range electrostatic forces, hydrogen constrains and \qty{2}{\femto\second} simulation steps were used.

For equilibration, first an energy minimization run with a maximum force of \qty{10}{\kilo\joule \per\mol\per\nano\meter} was done. Then two \qty{100}{\pico\second} simulation runs of an NVT and NPT ensemble were carried out, respectively.

Production simulations were done for \qty{200}{\nano\second} (apart from three systems for \qty{300}{\nano\second}) and with 32 replicas for each system with a geometric temperature distribution between \qty{300}{\kelvin} and \qty{500}{\kelvin}. The full list of simulation parameters can be found in the Appendix \ref{appendix:simulation_params}.

\subsection{Post-Processing} \label{subsec:post-processing}

Adding further derived quantities to the data set, MDAnalysis (\cite{MDAnalysis1,MDAnalysis2}) was used to calculate the root-mean-square deviation (RMSD), root-mean-square fluctuation (RMSF) and radius of gyration. For the dot bracket annotation of base pairs and the sugar pucker angle, MDtraj (\cite{MDTraj}) and barnaba (\cite{bottaro2019barnaba}) were used.

The forces of all higher temperature trajectories were additionally recalculated at the lowest temperature \qty{300}{\kelvin} with the \texttt{rerun} function from GROMACS.

\section{Data Records} \label{sec:data_records}
mdRNAcat is available on Hugging Face at
\url{https://huggingface.co/datasets/HelmholtzAI-FZJ/mdRNAcat}
(\cite{dorn2026}), with access via direct download or the
Hugging Face API. For each system in the dataset there are two files in the Hierarchical Data Format version 5 (HDF5), both labeled with the respective PDB ID of the system. The first contains the coordinates and forces of the lowest temperature replica with \qty{1}{\pico \second} time steps. The second file, labeled with \texttt{replica}, is only saved in with \qty{100}{\pico \second} time steps, but contains all replicas, replica exchange meta-data and additional derived quantities.

The organization within the HDF5 files is kept as similar as possible to the formatting used in the mdCATH dataset (\cite{mirarchi2024mdcath}) to ensure compatibility of code. A detailed list of the saved properties and values can be found in Table \ref{table:datafields}. Each file has the PDB ID as the root group, which contains for each atom in the system \texttt{chain}, \texttt{element}, \texttt{resid}, \texttt{resname} and its atom number \texttt{z}. Additionally the number of atoms, residues and replicas, as well as the temperature minimum and maximum are saved, along with the \texttt{.pdb} file and the GROMACS topology file. The \texttt{replica} files contain the \texttt{repex} group with replica exchange parameters such as the energy per replica, exchange attempts, successful exchanges and temperature ladder. For each replica at temperature \texttt{<T>/} (so for the non-\texttt{replica} file only the lowest replica), there is a group containing the box parameters, temperature, number of frames, stride as well as the coordinates and forces. In the \texttt{replica} files, these groups also contain derived analytical values such as radius of gyration, root-mean-square deviation and fluctuation (RMSD, RMSF), base-pair dot brackets, sugar pucker angle and the forces scaled to the lowest temperature. The latter allows additional lowest-temperature-frames with physically correct forces. However, the distribution in the conformation space of these frames is not thermodynamically correct, which has to be taken account depending on the intended usage.

As of the publication date, mdRNAcat contains 48 different RNA systems with a total simulation time of \qty{316.8}{\micro \second} and a total file size of 699\,GB (Table \ref{table:dataset_stats}).

\begin{table}[t]
\caption[The hierarchical organization of mdRNAcat available as HDF5 files.]{mdRNAcat is made available as HDF5 files with an hierarchical organization as shown in this table. For each system in the data set, there are 2 files both labeled with the respective PDB ID. The first contains only the lowest trajectory with \qty{1}{\pico \second} time steps. The second file contains all replicas and additional derived metrics indicated with a \colorbox{replonly}{blue background} in the table with \qty{100}{\pico \second} time steps. Key: \texttt{<T>}, REST2 effective temperature (K); $N$, number of atoms; $R$, number of residues; $F$, trajectory length in frames ($=\,$\texttt{frameStridePs}); $M$, number of replicas; $K$, energy terms; $S$, energy samples; $X$, exchange records; $P$, log segments.\footnotemark[2]}
\label{table:datafields}
\centering
\small
\setlength{\tabcolsep}{5pt}
\begin{tabular}{@{}l l l l l@{}}
\toprule
\textbf{Field} & \textbf{Size} & \textbf{Type} & \textbf{Unit} & \textbf{Description} \\
\midrule
\texttt{<pdbid>/}                    &                     &         &              &  \\
\quad\texttt{chain}                  & $N$                 & string  &              & Chain ID \\
\quad\texttt{element}                & $N$                 & string  &              & Chemical element \\
\quad\texttt{resid}                  & $N$                 & integer &              & Residue number \\
\quad\texttt{resname}                & $N$                 & string  &              & Residue name \\
\quad\texttt{z}                      & $N$                 & integer &              & Atomic number \\
\quad\texttt{pdb}                    & $1$                 & string  &              & RNA reference PDB \\
\quad\texttt{top}                    & $1$                 & string  &              & GROMACS topology \\
\quad\texttt{.numProteinAtoms}       & $1$                 & integer &              & Number of (RNA) atoms $N$ (attribute) \\
\quad\texttt{.numResidues}           & $1$                 & integer &              & Number of residues $R$ (attribute) \\
\quad\texttt{.numReplicas}           & $1$                 & integer &              & Number of replicas $M$ (attribute) \\
\quad\texttt{.rest2TempMin/Max}      & $1$                 & float   & K            & REST2 ladder end points (attributes) \\
\quad\texttt{<T>/}                   &                     &         &              &  \\
\quad\quad\texttt{0/}                &                     &         &              &  \\
\quad\quad\quad\texttt{coords}                        & $F\times N\times 3$ & float   & \AA          & Atom coordinates \\
\quad\quad\quad\texttt{forces}                        & $F\times N\times 3$ & float   & kcal/mol/\AA & Forces \\
    \rowcolor{replonly}
    \quad\quad\quad\texttt{scaledForces}              & $F\times N\times 3$ & float   & kcal/mol/\AA & $\lambda{=}1$ reference forces \\
    \rowcolor{replonly}
    \quad\quad\quad\texttt{gyrationRadius}            & $F$                 & double  & nm           & Gyration radius \\
    \rowcolor{replonly}
    \quad\quad\quad\texttt{rmsd}                      & $F$                 & float   & nm           & RMSD w.r.t. first frame \\
    \rowcolor{replonly}
    \quad\quad\quad\texttt{rmsf}                      & $R$                 & float   & nm           & Per-residue RMSF \\
    \rowcolor{replonly}
    \quad\quad\quad\texttt{secondaryStructure}        & $F\times R$         & string  &              & Base-pairing dot-bracket \\
    \rowcolor{replonly}
    \quad\quad\quad\texttt{sugarPucker}               & $F\times R$         & float   & degree       & Ribose pseudorotation phase \\
\quad\quad\quad\texttt{box}                           & $3\times 3$         & float   & nm           & Simulation unit cell \\
\quad\quad\quad\texttt{.temperature}                  & $1$                 & float   & K            & Effective temperature (attribute) \\
\quad\quad\quad\texttt{.numFrames}                    & $1$                 & integer &              & Number of frames $F$ (attribute) \\
\quad\quad\quad\texttt{.frameStridePs}                & $1$                 & float   & ps           & Time between frames (attribute) \\
    \rowcolor{replonly}
    \quad\texttt{repex/}                    &                     &         &        &  \\
    \rowcolor{replonly}
    \quad\quad\texttt{terms}                & $K$                 & string  &        & Interaction-energy term names \\
    \rowcolor{replonly}
    \quad\quad\texttt{energy}               & $M\times S\times K$ & float   & kJ/mol & Per-state interaction energies \\
    \rowcolor{replonly}
    \quad\quad\texttt{energyTime}           & $M\times S$         & float   & ps     & Energy sample times \\
    \rowcolor{replonly}
    \quad\quad\texttt{energyNSamples}       & $M$                 & integer &        & Valid samples per state \\
    \rowcolor{replonly}
    \quad\quad\texttt{replicaIndex}         & $X\times M$         & integer &        & Demux occupancy (replica per state) \\
    \rowcolor{replonly}
    \quad\quad\texttt{replicaIndexTime}     & $X$                 & float   & ps     & Exchange-record times \\
    \rowcolor{replonly}
    \quad\quad\texttt{nAttempts}            & $P$                 & float   &        & Exchange attempts per log segment \\
    \rowcolor{replonly}
    \quad\quad\texttt{nExchanges}           & $P\times(M{-}1)$    & float   &        & Accepted exchanges per adjacent pair \\
    \rowcolor{replonly}
    \quad\quad\texttt{temps}                & $M$                 & float   & K      & Effective-temperature ladder \\
\midrule
    \rowcolor{replonly}
    \multicolumn{5}{@{}l}{\quad\texttt{<T'>/}, \dots\quad\emph{remaining replicas}} \\
\bottomrule
\end{tabular}
\end{table}
\footnotetext[2]{A comment on the units used in mdRNAcat can be found in the Appendix \protect\ref{appendix:units}}
\addtocounter{footnote}{1}

\section{Analysis and Validation} \label{sec:validation}
\subsection{Sequence and Structural-Motif Space Coverage}\label{subsec:seq-val}
\begin{figure}
    \centering
    \includegraphics[width=1.2\textwidth]{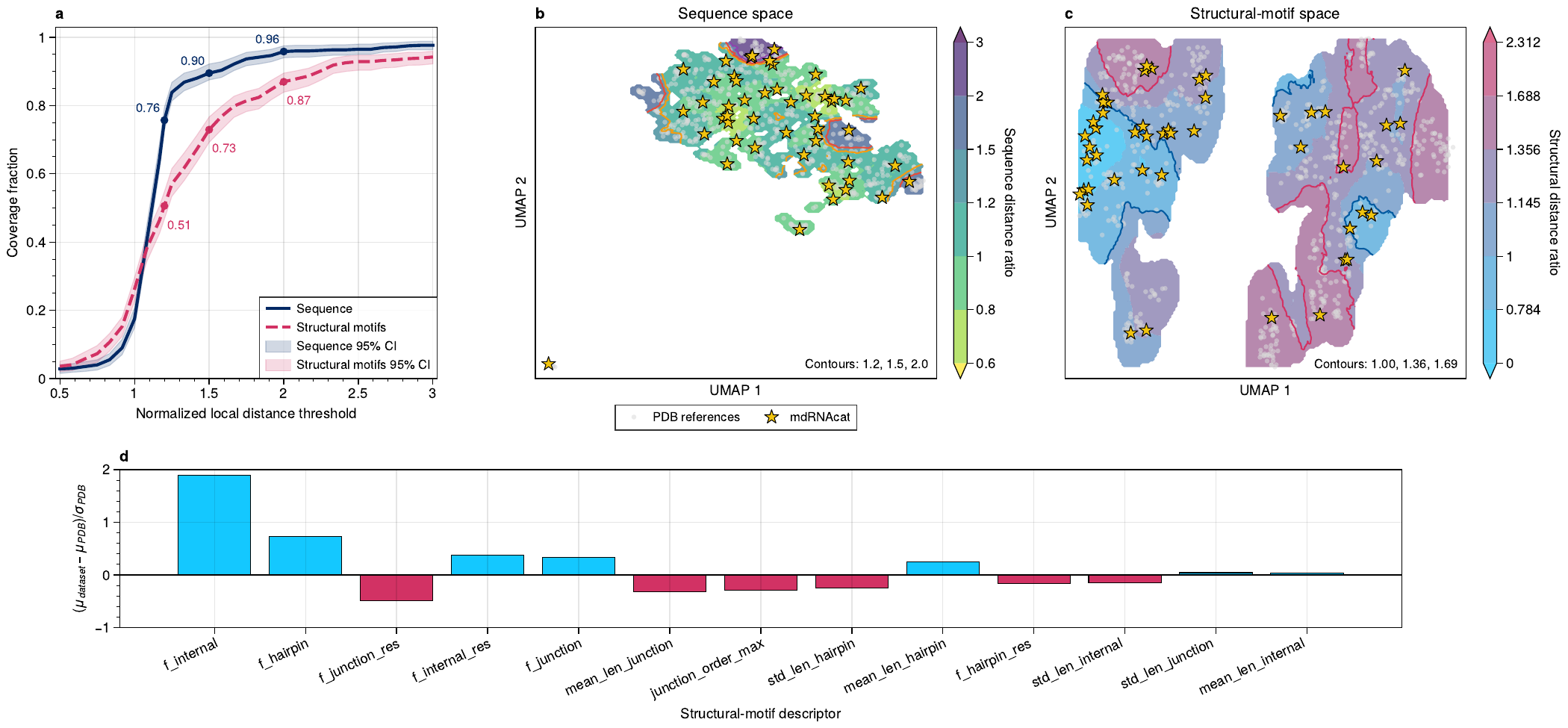}
    \caption{
        Sequence and structural-motif coverage of mdRNAcat against filtered X-ray PDB reference sets.
        (\textbf{a}) Coverage versus locally normalized nearest-representative distance.
        (\textbf{b},\textbf{c}) UMAP visualizations \cite{mcinnes2018_umap,mcinnes2018_JOSS}; stars denote mdRNAcat representatives and colors show smoothed distance ratios on separate scales, with structural values clipped for display.
        (\textbf{d}) Mean structural-descriptor differences in units of the reference-set standard deviation. Coverage is calculated before UMAP projection.
    }
    \label{plot:seq-val}
\end{figure}

We assessed mdRNAcat against filtered X-ray PDB reference sets released by 25 September 2026, reflecting the experimental method used for initial dataset selection (Figure~\ref{plot:seq-val}). All 48 mdRNAcat systems were included as representatives, irrespective of experimental method. Coverage was evaluated using Euclidean distances in SVD-reduced 3-mer TF--IDF space (\cite{moussa2018_tf_idf}) and in standardized space of 13 RNA 3D Hub structural-motif descriptors (\cite{petrov2013_rna3dhub}).

Distances to the nearest representative were normalized by the median distance to ten non-self neighbors (Appendix~\ref{appendix:coverage_metric}). Excluding mdRNAcat from evaluation, coverage at 1.5 local-neighbor scales reached \qty{89.4}{\percent} for sequences and \qty{72.9}{\percent} for structural motifs. The latter characterizes deposited structures, not the conformational ensembles sampled by REST2.

\subsection{Enhanced Sampling Validation} \label{subsec:sampling-val}

\begin{figure}
    \centering
    \includegraphics[width=1\textwidth]{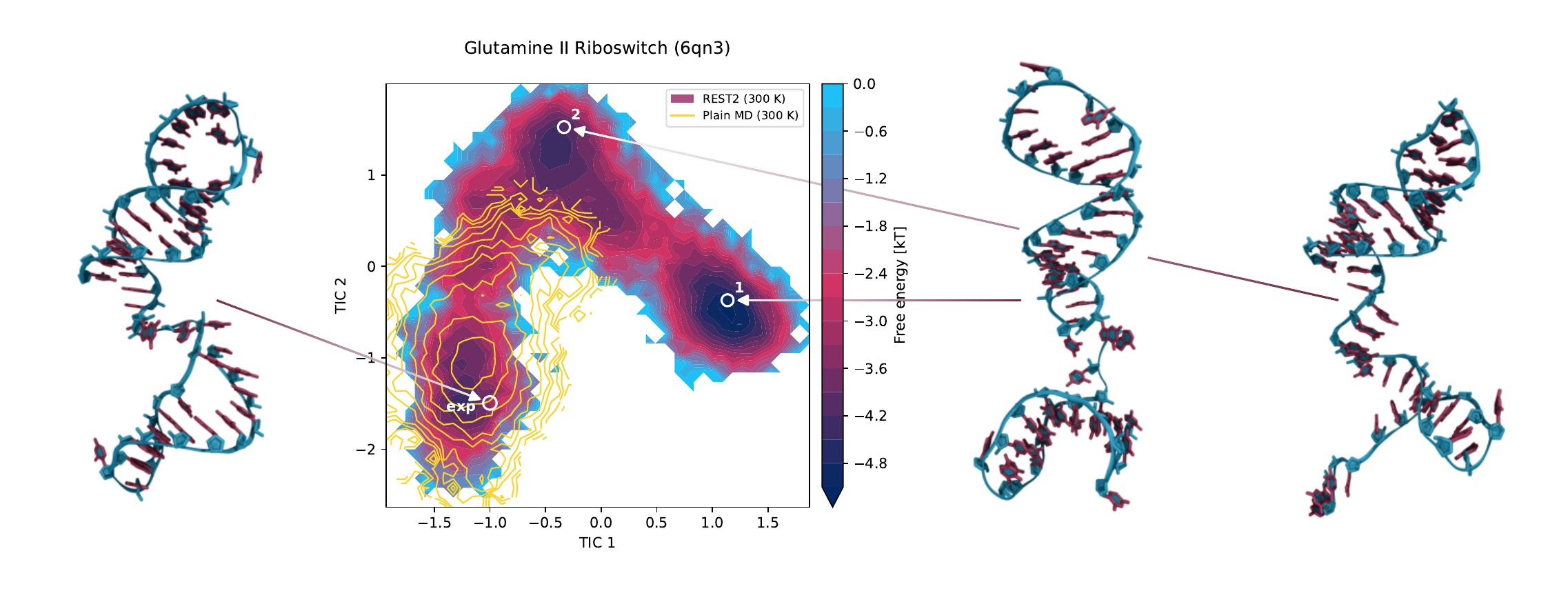}
    \caption{Free energy landscape (FEL) of the lowest temperature (\qty{300}{\kelvin}) trajectory of the 6qn3 system, the axes are determined with time-lagged independent component analysis (TICA) (\cite{TICA}), marked with the experimentally resolved structure (exp) and the two lowest minima (1 and 2); experimentally resolved structure is displayed to the left and structures from the two minima to the right; in yellow FEL of a plain single replica \qty{300}{\kelvin} simulation of 6qn3 (not part of mdRNAcat) mapped to the same TICA coordinates.}
    \label{plot:tica_rest2}
\end{figure}

To validate the methodological use of REST2 enhanced sampling of RNA conformational states, we compare the free energy landscapes (FELs) (\cite{bussi2020metadynamics,rohrdanz2013discovering}) obtained through sampling both plain MD and REST2 trajectories of an illustrative system, the Glutamine II Riboswitch (PDB ID: 6qn3) (Figure \ref{plot:tica_rest2}). The REST2 landscape is constructed only from the lowest temperature trajectory, which is in thermodynamic equilibrium. Compressing the $3 \cdot N_{atoms}$-dimensional conformation space down to two dimensions and enabling planar visualization requires targeted dimensionality reduction techniques. The landscapes are constructed using time-lagged independent component analysis (TICA) (\cite{TICA}), which gives the axis of slowest dynamical motion along the trajectory. 
The plain MD simulation is not included in the mdRNAcat dataset.
It has the same pre-processing and simulation parameters as the REST2 simulations with the same simulation length of \qty{200}{\nano\second}, apart from sampling only a single trajectory at \qty{300}{\kelvin} (which is equivalent to the lowest temperature of the REST2 simulations) without any replica exchange.

For the REST2 simulation, one can see that the trajectory successfully samples outside of the experimentally resolved state. There are three distinct energy minima which get sampled more often than the rest of the conformation space, one of which contains the experimentally resolved structure. The plain MD simulation only finds minima of the experimentally resolved state where it started and fails to find the other important ensemble states. In total only \qty{47.6}{\percent} of the area sampled by REST2 is also sampled in the plain MD. This result sufficiently demonstrates the enhanced sampling offered by state-of-the-art algorithms such as REST2 which we used for this dataset.

\subsection{Replica Exchange Validation} \label{subsec:repex-val}

To ensure that the replica exchange algorithm performed as intended, we monitored the exchange statistics and characteristics of each system (Figure \ref{plot:repex} as an illustrative system, all other systems can be found in the Appendix \ref{appendix:overlap_plots}). Replica exchange requires exchange rates around 20\,\% (\cite{kone2005selection,rathore2005optimal}) are favorable, which we checked in (a). For exchanges to happen, there needs to be an adequate overlap in the energy distribution of the different thermodynamic states (which can be seen in (b)). Lastly, subplot (c) does not track the thermodynamic state, but the energy distribution of the frames of a single replica that gets exchanged to higher and lower temperatures. This demonstrates that they are not stuck in certain temperatures. A statistical summary over all systems can be found in Table \ref{table:repex}.

\begin{figure}
    \centering
    \includegraphics[width=0.9\textwidth]{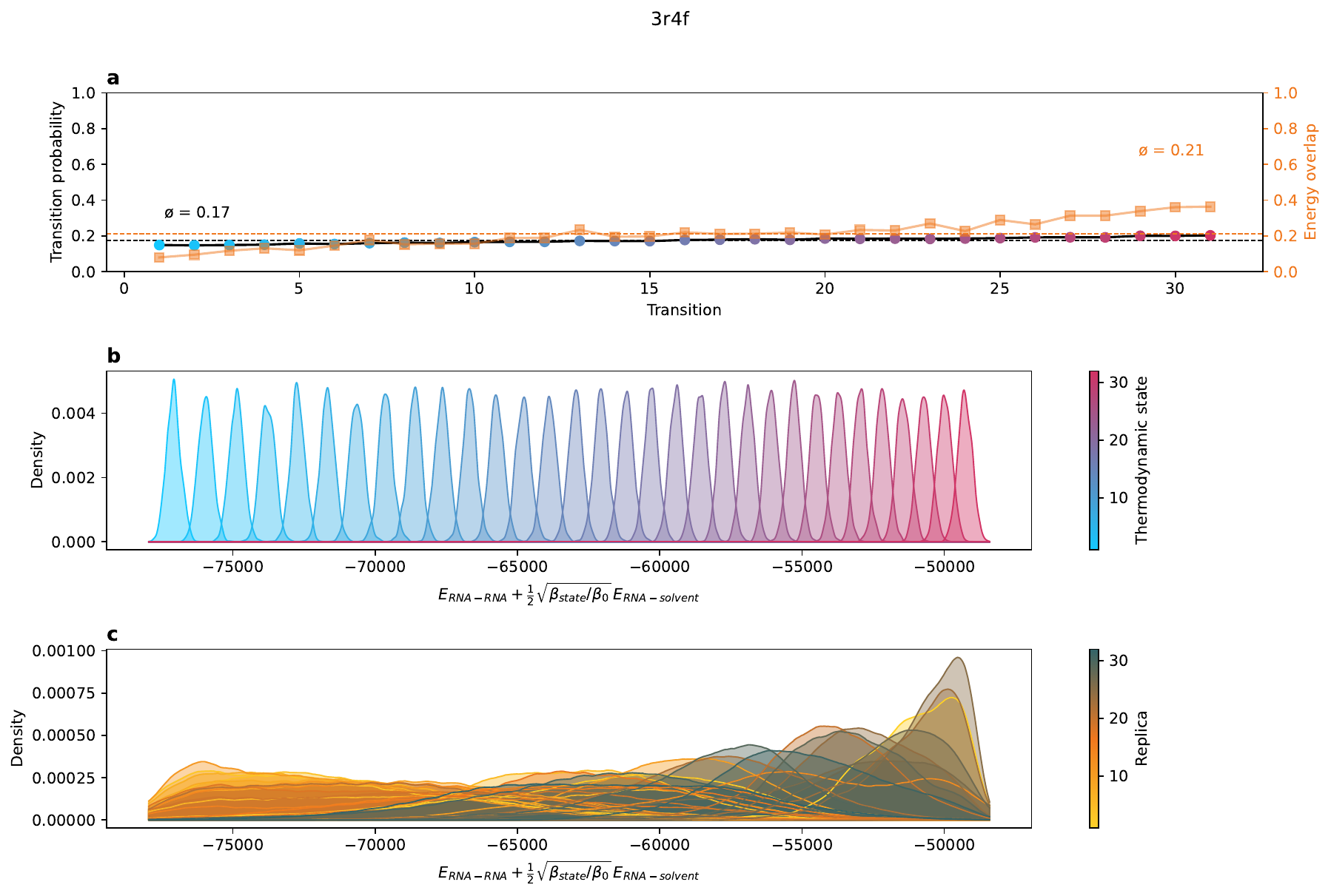}
    \caption{(\textbf{a}) Transition probability between two thermodynamic states and energy overlap between two thermodynamic states against the 31 transitions of the states for the 3r4f system, with dotted lines indicating the average over all transitions. (\textbf{b}) Density of the thermodynamic states against the energy of the same system. (\textbf{c}) Density of the replicas against the energy of the same system.}
    \label{plot:repex}
\end{figure}

\begin{table}[t]
\caption{Replica exchange statistics averaged over all 48 systems with standard deviation between systems.}
\label{table:repex}
\centering
\small
\begin{tabular}{lll}
\toprule
Avg. transition probability & $0.183$ & $\pm 0.065$ \\
Avg. energy overlap & $0.269$ & $\pm 0.070$ \\
\bottomrule
\end{tabular}
\end{table}

\begin{figure}
    \centering
    \includegraphics[width=0.8\textwidth]{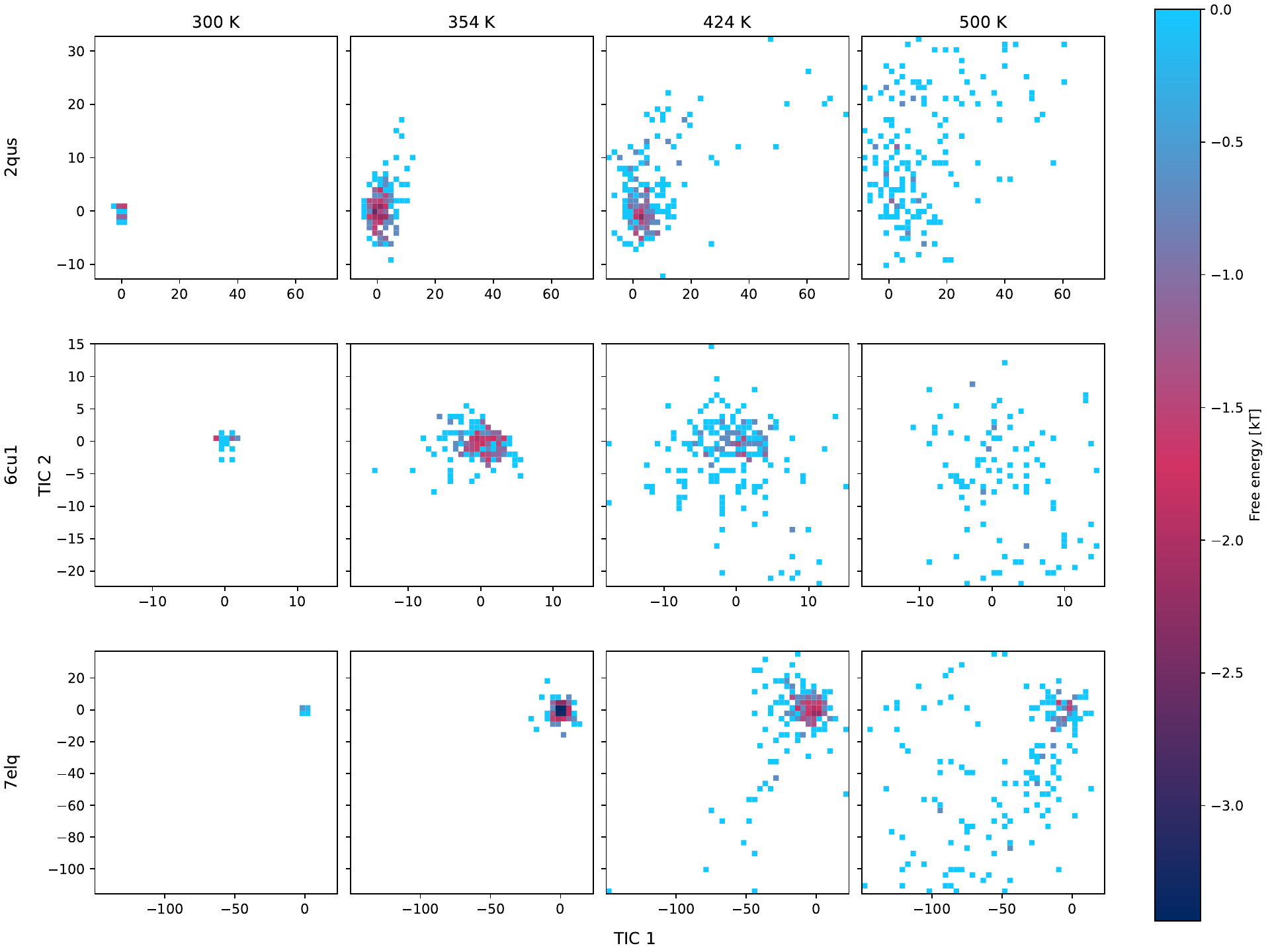}
    \caption{Free energy landscape (FEL) of the 2qus, 6cu1 and 7elq systems, the axes are determined with time-lagged independent component analysis (TICA) (\cite{TICA}) at \qty{300}{\kelvin}, \qty{354}{\kelvin}, \qty{424}{\kelvin} and \qty{500}{\kelvin}. Note that the axes are capped at the extension of the \qty{424}{\kelvin}-plot, so the \qty{500}{\kelvin}-plot has values outside of that range not displayed. The conformation space gets explored more with higher temperature.}
    \label{plot:replica_tica}
\end{figure}

In higher energy replicas, one expects a broader sampling of the conformation space. We plotted the illustrative FEL for 3 different systems at 4 different temperatures (Figure \ref{plot:replica_tica}). In all systems, the conformation space is explored more widely with rising temperature. This confirms the validity of the temperature scaling.

\subsection{Base Pairing and Backbone Torsion} \label{subsec:base-sugar-val}

\begin{figure}
    \centering
    \includegraphics[width=1\textwidth]{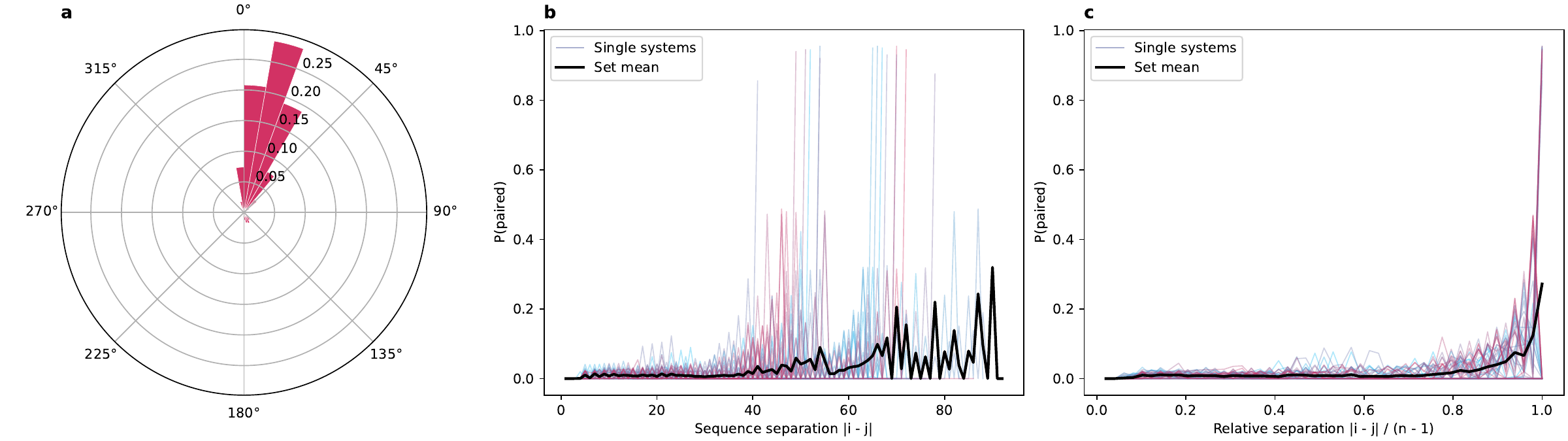}
    \caption{(\textbf{a}) Wheel histogram over the sugar-pucker angle of all residues of all systems; north (\qty{0}{\degree}) indicates C3'-endo, so A-form helical pucker; south (\qty{180}{\degree}) indicates C2'-endo, so loops and bulges. (\textbf{b}) Distribution of the residue pair likelihood against the sequence separation between the two paired residue for all systems in the data set; thick is the set mean. (\textbf{c}) Same as (\textbf{b}) but the x-axis normalized to 1.} 
    \label{plot:base_sugar}
\end{figure}

In this section we wanted to validate the structures of the system. The first metric looked at is the angle of the sugar pucker in the backbone. This indicates the angle torsion and thus the secondary structure of the RNA (\cite{altona1972conformational,saenger1984furanose}). In Figure \ref{plot:base_sugar} (a), the percentage of each torsion angle is plotted in a wheel plot. A north facing angle means C3'-endo conformation and an A-form helix, which most residues and frames in the dataset represent. There is also a small fraction facing south indicating C2'-endo conformation in loops and bulges. 

Additionally, we investigated the base pairings. In Figure \ref{plot:base_sugar} (b) and (c), the averaged probability for each sequence separation between two paired bases is plotted. As the systems are not of the same length, the tail of the distribution in (b) is dominated by a few larger systems. In (c), the separation distance is normed to 1, so that all systems in the dataset are on the axis. This reveals that the pair probability grows with the separation distance. This is due to the double helical conformation, pairing both ends of a sequence, while the middle part forms a loop, confirming the results of the backbone torsion analysis.

\begin{figure}
    \centering
    \includegraphics[width=1\textwidth]{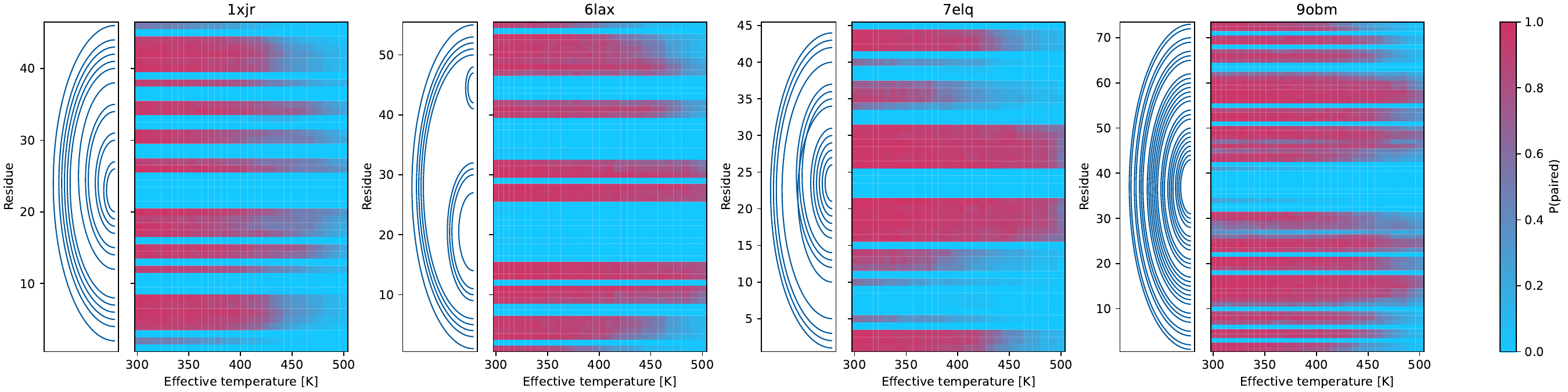}
    \caption{Probability of a paired residue in residue and temperature dimensions for the 1xjr, 6lax, 7elq and 9obm systems. On the left of each plot is a bow plot indicating the paired residues in the experimentally resolved structure.}
    \label{plot:melting}
\end{figure}

Lastly, we expect base pairs to break in higher energy states, for that we plotted the pair probability per residue and per temperature (Figure \ref{plot:melting}). On the left of each plot, bow plots also indicate which bases are paired in the experimentally resolved structure. Again one can observe that the ends are paired and the center builds a loop. With higher temperature the pairs are breaking, usually starting with the ends, while the center of the RNA stays stable for higher energies. This is expected behavior for RNA and further confirms the validity of our simulations.


\subsection*{AI use statement}

In this work, we used generative AI tools for writing part of the code used for post-processing and plotting.
We have not used generative AI tools for the conception of the dataset, any pre-processing or the simulations and the generated data themselves.
The rest of the required disclosure tasks such as mathematical claims or proofs are not applicable to this work.
Additionally, we used generative AI tools for searching relevant literature and other works. 
We have reviewed all AI-assisted work. LLM-generated code was verified and tested for correctness by 2 authors. We take responsibility for the final content of this work, including text, claims or artifacts produced with the aid of generative AI.

\subsection*{Reproducibility statement}

To ensure the reproducibility of our work, we included all \texttt{.pdb} and topology files in the dataset itself. Additionally detailed record of all pre-processing and simulation parameters can be found in the Appendix.

The code used to create the validation done in this paper can be found under: \url{https://github.com/SDLAML/mdRNAcat/}

\subsubsection*{Author Contributions}
Anton E. Dorn (Conceptualization, Methodology, Simulation, Validation, Data curation, Writing--original draft), Emile de Bruyn (Conceptualization, Methodology, Validation, Data curation, Writing--original draft), Fabrice von der Lehr (Conceptualization, Writing--review \& editing), Philipp Knechtges (Conceptualization, Funding acquisition), Stefan Kesselheim (Conceptualization, Funding acquisition, Writing--review \& editing), Alexander Schug (Conceptualization, Funding acquisition, Writing--review \& editing)

\subsubsection*{Acknowledgments}
The authors gratefully acknowledge computing time on the supercomputer JURECA (\cite{JURECA}) at Forschungszentrum Jülich under grant no. 65978 (EATS-RNA). 
We furthermore gratefully acknowledge the Gauss Centre for Supercomputing e.V. and the John von Neumann Institute for Computing (NIC) for computing time on JUPITER at Jülich Supercomputing Centre, including resources awarded through the NIC Excellence project PROFOUND. 
A.S. acknowledges support from HIDSS4Health and from the Helmholtz Association's Initiative and Networking Fund through the Helmholtz AI platform and ZebraTwin. 
E.d.B. and S.K. acknowledge support by the German Federal Ministry for Research, Technology and Space (BMFTR) under grant number 01IS24076B (FAIME). 
A.E.D., E.d.B., S.K. and A.S. acknowledge support from the Helmholtz Foundation Model Initiative (HFMI) through PROFOUND and Virtual Cell. 
All authors acknowledge support from Helmholtz AI through the MADRNA project.
The funding bodies acknowledged above were not involved in the study design, data collection, analysis, decision to publish, or preparation of the manuscript.

\bibliography{iclr2027_conference}

@article{vander2024atlas,
  title={ATLAS: protein flexibility description from atomistic molecular dynamics simulations},
  author={Vander Meersche, Yann and Cretin, Gabriel and Gheeraert, Aria and Gelly, Jean-Christophe and Galochkina, Tatiana},
  journal={Nucleic acids research},
  volume={52},
  number={D1},
  pages={D384--D392},
  year={2024},
  publisher={Oxford University Press}
}

@article{bussi2020metadynamics,
  author  = {Bussi, Giovanni and Laio, Alessandro},
  title   = {Using Metadynamics to Explore Complex Free-Energy Landscapes},
  journal = {Nature Reviews Physics},
  year    = {2020},
  volume  = {2},
  number  = {4},
  pages   = {200--212},
  doi     = {10.1038/s42254-020-0153-0}
}

@article{kouba2023machine,
  title={Machine learning-guided protein engineering},
  author={Kouba, Petr and Kohout, Pavel and Haddadi, Faraneh and Bushuiev, Anton and Samusevich, Raman and Sedlar, Jiri and Damborsky, Jiri and Pluskal, Tomas and Sivic, Josef and Mazurenko, Stanislav},
  journal={ACS catalysis},
  volume={13},
  number={21},
  pages={13863},
  year={2023}
}

@article{hospital2016bignasim,
  author  = {Hospital, Adam and Andrio, Pau and Cugnasco, Carles and Codo, Laia and Becerra, Yolanda and Dans, Pablo D. and Battistini, Federica and Torres, Javier and Go{\~n}i, Ramon and Orozco, Modesto and Gelp{\'i}, Josep L.},
  title   = {BIGNASim: a NoSQL database structure and analysis portal for nucleic acids simulation data},
  journal = {Nucleic Acids Research},
  volume  = {44},
  number  = {D1},
  pages   = {D272--D278},
  year    = {2016},
  doi     = {10.1093/nar/gkv1301}
}

@article{mirarchi2024mdcath,
  title={mdCATH: A large-scale MD dataset for data-driven computational biophysics},
  author={Mirarchi, Antonio and Giorgino, Toni and De Fabritiis, Gianni},
  journal={Scientific Data},
  volume={11},
  number={1},
  pages={1299},
  year={2024},
  publisher={Nature Publishing Group UK London}
}

@article{alphafold3,
  author =       {Josh Abramson and Jonas Adler and Jack Dunger and Richard
                  Evans and Tim Green and Alexander Pritzel and Olaf Ronneberger
                  and Lindsay Willmore and Andrew J. Ballard and Joshua Bambrick
                  and Sebastian W. Bodenstein and David A. Evans and Chia-Chun
                  Hung and Michael O'Neill and David Reiman and Kathryn
                  Tunyasuvunakool and Zachary Wu and Akvilė Žemgulytė and Eirini
                  Arvaniti and Charles Beattie and Ottavia Bertolli and Alex
                  Bridgland and Alexey Cherepanov and Miles Congreve and
                  Alexander I. Cowen-Rivers and Andrew Cowie and Michael
                  Figurnov and Fabian B. Fuchs and Hannah Gladman and Rishub
                  Jain and Yousuf A. Khan and Caroline M. R. Low and Kuba Perlin
                  and Anna Potapenko and Pascal Savy and Sukhdeep Singh and
                  Adrian Stecula and Ashok Thillaisundaram and Catherine Tong
                  and Sergei Yakneen and Ellen D. Zhong and Michal Zielinski and
                  Augustin Ž{\'i}dek and Victor Bapst and Pushmeet Kohli and Max
                  Jaderberg and Demis Hassabis and John M. Jumper},
  title =        {Accurate Structure Prediction of Biomolecular Interactions
                  With Alphafold 3},
  journal =      {Nature},
  volume =       630,
  number =       8016,
  pages =        {493-500},
  year =         2024,
  doi =          {10.1038/s41586-024-07487-w},
  url =          {http://dx.doi.org/10.1038/s41586-024-07487-w},
  DATE_ADDED =   {Thu Sep 17 14:22:03 2026},
}

@article{nucleoseeker,
  author =       {Utkarsh Upadhyay and Fabrizio Pucci and Julian Herold and
                  Alexander Schug},
  title =        {Nucleoseeker-Precision Filtering of Rna Databases To Curate
                  High-Quality Datasets},
  journal =      {NAR Genomics and Bioinformatics},
  volume =       7,
  number =       1,
  pages =        {lqaf021},
  year =         2025,
  doi =          {10.1093/nargab/lqaf021},
  url =          {http://dx.doi.org/10.1093/nargab/lqaf021},
  DATE_ADDED =   {Thu Feb 5 10:25:46 2026},
}

@article{rest2,
  author =       {Lingle Wang and Richard A. Friesner and B. J. Berne},
  title =        {Replica Exchange With Solute Scaling: a More Efficient Version
                  of Replica Exchange With Solute Tempering (REST2)},
  journal =      {The Journal of Physical Chemistry B},
  volume =       115,
  number =       30,
  pages =        {9431-9438},
  year =         2011,
  doi =          {10.1021/jp204407d},
  url =          {https://doi.org/10.1021/jp204407d},
  DATE_ADDED =   {Wed Mar 3 09:58:52 2021},
}

@article{PLUMED3,
  author =       {The PLUMED consortium},
  title =        {Promoting Transparency and Reproducibility in Enhanced
                  Molecular Simulations},
  journal =      {Nature Methods},
  volume =       16,
  number =       8,
  pages =        {670-673},
  year =         2019,
  doi =          {10.1038/s41592-019-0506-8},
  url =          {https://doi.org/10.1038/s41592-019-0506-8},
  DATE_ADDED =   {Mon Aug 23 08:48:33 2021},
}

@article{PLUMED2,
  author =       {Gareth A. Tribello and Massimiliano Bonomi and Davide
                  Branduardi and Carlo Camilloni and Giovanni Bussi},
  title =        {Plumed 2: New Feathers for an Old Bird},
  journal =      {Computer Physics Communications},
  volume =       185,
  number =       2,
  pages =        {604-613},
  year =         2014,
  doi =          {10.1016/j.cpc.2013.09.018},
  url =          {https://doi.org/10.1016/j.cpc.2013.09.018},
  DATE_ADDED =   {Mon Aug 23 08:49:03 2021},
}

@article{PLUMED1,
  author =       {Massimiliano Bonomi and Davide Branduardi and Giovanni Bussi
                  and Carlo Camilloni and Davide Provasi and Paolo Raiteri and
                  Davide Donadio and Fabrizio Marinelli and Fabio Pietrucci and
                  Ricardo A. Broglia and Michele Parrinello},
  title =        {Plumed: a Portable Plugin for Free-Energy Calculations With
                  Molecular Dynamics},
  journal =      {Computer Physics Communications},
  volume =       180,
  number =       10,
  pages =        {1961-1972},
  year =         2009,
  doi =          {10.1016/j.cpc.2009.05.011},
  url =          {https://doi.org/10.1016/j.cpc.2009.05.011},
  DATE_ADDED =   {Mon Aug 23 08:50:45 2021},
}

@article{MDTraj,
  author =       {Robert T. McGibbon and Kyle A. Beauchamp and Matthew P.
                  Harrigan and Christoph Klein and Jason M. Swails and Carlos X.
                  Hern{\'a}ndez and Christian R. Schwantes and Lee-Ping Wang and
                  Thomas J. Lane and Vijay S. Pande},
  title =        {Mdtraj: a Modern Open Library for the Analysis of Molecular
                  Dynamics Trajectories},
  journal =      {Biophysical Journal},
  volume =       109,
  number =       8,
  pages =        {1528-1532},
  year =         2015,
  doi =          {10.1016/j.bpj.2015.08.015},
  url =          {https://doi.org/10.1016/j.bpj.2015.08.015},
  DATE_ADDED =   {Tue Mar 9 16:49:57 2021},
}

@inproceedings{MDAnalysis2,
  author =       {Richard Gowers and Max Linke and Jonathan Barnoud and Tyler
                  Reddy and Manuel Melo and Sean Seyler and Jan Domański and
                  David Dotson and S{\'e}bastien Buchoux and Ian Kenney and
                  Oliver Beckstein},
  title =        {MDAnalysis: A Python Package for the Rapid Analysis of
                  Molecular Dynamics Simulations},
  booktitle =    {Proceedings of the 15th Python in Science Conference},
  year =         2016,
  pages =        {nil},
  doi =          {10.25080/majora-629e541a-00e},
  url =          {https://doi.org/10.25080/majora-629e541a-00e},
  DATE_ADDED =   {Tue Mar 9 16:53:13 2021},
}

@article{MDAnalysis1,
  author =       {Naveen Michaud-Agrawal and Elizabeth J. Denning and Thomas B.
                  Woolf and Oliver Beckstein},
  title =        {Mdanalysis: a Toolkit for the Analysis of Molecular Dynamics
                  Simulations},
  journal =      {Journal of Computational Chemistry},
  volume =       32,
  number =       10,
  pages =        {2319-2327},
  year =         2011,
  doi =          {10.1002/jcc.21787},
  url =          {https://doi.org/10.1002/jcc.21787},
  DATE_ADDED =   {Tue Mar 9 16:54:03 2021},
}

@article{TICA,
  author =       {Guillermo P{\'e}rez-Hern{\'a}ndez and Fabian Paul and Toni
                  Giorgino and Gianni De Fabritiis and Frank No{\'e}},
  title =        {Identification of Slow Molecular Order Parameters for Markov
                  Model Construction},
  journal =      {The Journal of Chemical Physics},
  volume =       139,
  number =       1,
  pages =        015102,
  year =         2013,
  doi =          {10.1063/1.4811489},
  url =          {http://dx.doi.org/10.1063/1.4811489},
  DATE_ADDED =   {Sun Apr 9 13:29:42 2023},
}

@article{RNA_CG_FF_review,
  author =       {Wenfei Li and Shi-Jie Chen},
  title =        {Building Rna Coarse-Grained Force Fields: Design Principles
                  and Training Strategies},
  journal =      {Biophysical Journal},
  volume =       125,
  number =       18,
  pages =        {4898-4912},
  year =         2026,
  doi =          {10.1016/j.bpj.2026.03.041},
  url =          {http://dx.doi.org/10.1016/j.bpj.2026.03.041},
  DATE_ADDED =   {Fri Sep 18 17:11:45 2026},
}

@article{hollingsworth2018molecular,
  title={Molecular dynamics simulation for all},
  author={Hollingsworth, Scott A and Dror, Ron O},
  journal={Neuron},
  volume={99},
  number={6},
  pages={1129--1143},
  year={2018},
  publisher={Elsevier}
}

@article{kovacs2025mace,
  title={Mace-off: Short-range transferable machine learning force fields for organic molecules},
  author={Kov{\'a}cs, D{\'a}vid P{\'e}ter and Moore, J Harry and Browning, Nicholas J and Batatia, Ilyes and Horton, Joshua T and Pu, Yixuan and Kapil, Venkat and Witt, William C and Magdau, Ioan-Bogdan and Cole, Daniel J and others},
  journal={Journal of the American Chemical Society},
  volume={147},
  number={21},
  pages={17598--17611},
  year={2025},
  publisher={ACS Publications}
}

@article{upadhyay2026sequence,
  title={From sequence to structure: A comprehensive review of deep learning models for RNA structure prediction},
  author={Upadhyay, Utkarsh and Dorn, Anton and Faber, Christian and Schug, Alexander},
  journal={Current Opinion in Structural Biology},
  volume={97},
  pages={103216},
  year={2026},
  publisher={Elsevier}
}

@article{berman2000protein,
  title={The protein data bank},
  author={Berman, Helen M and Westbrook, John and Feng, Zukang and Gilliland, Gary and Bhat, Talapady N and Weissig, Helge and Shindyalov, Ilya N and Bourne, Philip E},
  journal={Nucleic acids research},
  volume={28},
  number={1},
  pages={235--242},
  year={2000},
  publisher={Oxford University Press}
}

@article{levine2025open,
  title={The open molecules 2025 (omol25) dataset, evaluations, and models},
  author={Levine, Daniel S and Shuaibi, Muhammed and Spotte-Smith, Evan Walter Clark and Taylor, Michael G and Hasyim, Muhammad R and Michel, Kyle and Batatia, Ilyes and Cs{\'a}nyi, G{\'a}bor and Dzamba, Misko and Eastman, Peter and others},
  journal={arXiv preprint arXiv:2505.08762},
  year={2025}
}

@article{eastman2024nutmeg,
  title={Nutmeg and spice: models and data for biomolecular machine learning},
  author={Eastman, Peter and Pritchard, Benjamin P and Chodera, John D and Markland, Thomas E},
  journal={Journal of chemical theory and computation},
  volume={20},
  number={19},
  pages={8583},
  year={2024}
}

@article{abrams2013enhanced,
  title={Enhanced sampling in molecular dynamics using metadynamics, replica-exchange, and temperature-acceleration},
  author={Abrams, Cameron and Bussi, Giovanni},
  journal={Entropy},
  volume={16},
  number={1},
  pages={163--199},
  year={2013},
  publisher={Molecular Diversity Preservation International (MDPI)}
}

@article{abraham2015gromacs,
  title={GROMACS: High performance molecular simulations through multi-level parallelism from laptops to supercomputers},
  author={Abraham, Mark James and Murtola, Teemu and Schulz, Roland and P{\'a}ll, Szil{\'a}rd and Smith, Jeremy C. and Hess, Berk and Lindahl, Erik},
  journal={SoftwareX},
  volume={1},
  pages={19--25},
  year={2015},
  publisher={Elsevier}
}

@article{case2025recent,
  title={Recent developments in Amber biomolecular simulations},
  author={Case, David A and Cerutti, David S and Cruzeiro, Vin{\'\i}cius Wilian D and Darden, Thomas A and Duke, Robert E and Ghazimirsaeed, Mahdieh and Giambasu, George M and Giese, Timothy J and Gotz, Andreas W and Harris, Julie A and others},
  journal={Journal of chemical information and modeling},
  volume={65},
  number={15},
  pages={7835--7843},
  year={2025},
  publisher={ACS Publications}
}

@article{perez2007refinement,
  title={Refinement of the AMBER force field for nucleic acids: improving the description of $\alpha$/$\gamma$ conformers},
  author={P{\'e}rez, Alberto and March{\'a}n, Iv{\'a}n and Svozil, Daniel and Sponer, Jiri and Cheatham, Thomas E and Laughton, Charles A and Orozco, Modesto},
  journal={Biophysical journal},
  volume={92},
  number={11},
  pages={3817--3829},
  year={2007},
  publisher={Elsevier}
}

@article{guy2012single,
  title={Single-stranded DNA within nanopores: conformational dynamics and implications for sequencing; a molecular dynamics simulation study},
  author={Guy, Andrew T and Piggot, Thomas J and Khalid, Syma},
  journal={Biophysical journal},
  volume={103},
  number={5},
  pages={1028--1036},
  year={2012},
  publisher={Elsevier}
}

@article{tan2018rna,
  title={RNA force field with accuracy comparable to state-of-the-art protein force fields},
  author={Tan, Dazhi and Piana, Stefano and Dirks, Robert M and Shaw, David E},
  journal={Proceedings of the National Academy of Sciences},
  volume={115},
  number={7},
  pages={E1346--E1355},
  year={2018},
  publisher={National Academy of Sciences}
}

@article{piana2015water,
  title={Water dispersion interactions strongly influence simulated structural properties of disordered protein states},
  author={Piana, Stefano and Donchev, Alexander G and Robustelli, Paul and Shaw, David E},
  journal={The journal of physical chemistry B},
  volume={119},
  number={16},
  pages={5113--5123},
  year={2015},
  publisher={ACS Publications}
}

@article{mackerell1998all,
  title={All-atom empirical potential for molecular modeling and dynamics studies of proteins},
  author={MacKerell Jr, Alex D and Bashford, Donald and Bellott, MLDR and Dunbrack Jr, Roland Leslie and Evanseck, Jeffrey D and Field, Martin J and Fischer, Stefan and Gao, Jiali and Guo, Houyang and Ha, Sookhee and others},
  journal={The journal of physical chemistry B},
  volume={102},
  number={18},
  pages={3586--3616},
  year={1998},
  publisher={ACS Publications}
}

@article{bottaro2019barnaba,
  title={Barnaba: software for analysis of nucleic acid structures and trajectories},
  author={Bottaro, Sandro and Bussi, Giovanni and Pinamonti, Giovanni and Rei{\ss}er, Sabine and Boomsma, Wouter and Lindorff-Larsen, Kresten},
  journal={Rna},
  volume={25},
  number={2},
  pages={219--231},
  year={2019},
  publisher={Cold Spring Harbor Laboratory Press}
}

@article{kone2005selection,
  title={Selection of temperature intervals for parallel-tempering simulations},
  author={Kone, Aminata and Kofke, David A},
  journal={The Journal of chemical physics},
  volume={122},
  number={20},
  year={2005},
  publisher={AIP Publishing}
}

@article{rathore2005optimal,
  title={Optimal allocation of replicas in parallel tempering simulations},
  author={Rathore, Nitin and Chopra, Manan and de Pablo, Juan J},
  journal={The Journal of chemical physics},
  volume={122},
  number={2},
  pages={024111},
  year={2005},
  publisher={American Institute of Physics}
}

@article{rohrdanz2013discovering,
  title={Discovering mountain passes via torchlight: Methods for the definition of reaction coordinates and pathways in complex macromolecular reactions},
  author={Rohrdanz, Mary A and Zheng, Wenwei and Clementi, Cecilia},
  journal={Annual review of physical chemistry},
  volume={64},
  number={1},
  pages={295--316},
  year={2013},
  publisher={Annual Reviews}
}

@article{altona1972conformational,
  title={Conformational analysis of the sugar ring in nucleosides and nucleotides. New description using the concept of pseudorotation},
  author={Altona, C. and Sundaralingam, M.},
  journal={Journal of the American Chemical Society},
  volume={94},
  number={23},
  pages={8205--8212},
  year={1972},
  publisher={ACS Publications}
}

@incollection{saenger1984furanose,
  author    = {Saenger, Wolfram},
  title     = {Structures and Conformational Properties of Bases,
               Furanose Sugars, and Phosphate Groups},
  booktitle = {Principles of Nucleic Acid Structure},
  chapter   = {4},
  series    = {Springer Advanced Texts in Chemistry},
  publisher = {Springer},
  address   = {New York},
  year      = {1984},
  isbn      = {978-0-387-90761-1},
  doi       = {10.1007/978-1-4612-5190-3}
}

@article{steinbrecher2012revised,
  title={Revised AMBER parameters for bioorganic phosphates},
  author={Steinbrecher, T and Latzer, J and Case, DA},
  journal={Journal of chemical theory and computation},
  volume={8},
  number={11},
  pages={4405},
  year={2012}
}

@article{moussa2018_tf_idf,
  author =       {Marmar Moussa and Ion I. Măndoiu},
  title =        {Single Cell Rna-Seq Data Clustering Using Tf-Idf Based
                  Methods},
  journal =      {BMC Genomics},
  volume =       19,
  number =       {S6},
  pages =        569,
  year =         2018,
  doi =          {10.1186/s12864-018-4922-4},
  url =          {http://dx.doi.org/10.1186/s12864-018-4922-4},
  DATE_ADDED =   {Thu Sep 24 10:55:38 2026},
}

@article{mcinnes2018_JOSS,
  author =       {Leland McInnes and John Healy and Nathaniel Saul and Lukas
                  Gro{\ss}berger},
  title =        {Umap: Uniform Manifold Approximation and Projection},
  journal =      {Journal of Open Source Software},
  volume =       3,
  number =       29,
  pages =        861,
  year =         2018,
  doi =          {10.21105/joss.00861},
  url =          {http://dx.doi.org/10.21105/joss.00861},
  DATE_ADDED =   {Thu Sep 24 11:07:20 2026},
}

@article{mcinnes2018_umap,
  author =       {McInnes, Leland and Healy, John and Melville, James},
  title =        {Umap: Uniform Manifold Approximation and Projection for
                  Dimension Reduction},
  journal =      {CoRR},
  year =         2018,
  url =          {http://arxiv.org/abs/1802.03426v3},
  archivePrefix ={arXiv},
  eprint =       {1802.03426v3},
  primaryClass = {stat.ML},
}

@article{petrov2013_rna3dhub,
  author =       {Anton I. Petrov and Craig L. Zirbel and Neocles B. Leontis},
  title =        {Automated Classification of Rna 3d Motifs and the Rna 3d Motif
                  Atlas},
  journal =      {RNA},
  volume =       19,
  number =       10,
  pages =        {1327-1340},
  year =         2013,
  doi =          {10.1261/rna.039438.113},
  url =          {http://dx.doi.org/10.1261/rna.039438.113},
  DATE_ADDED =   {Fri Sep 25 20:23:15 2026},
}

@misc{dorn2026,
	author       = { Anton E. Dorn and Emile de Bruyn and Fabrice von der Lehr and Philipp Knechtges and Stefan Kesselheim and Alexander Schug },
	title        = { mdRNAcat (Revision 0545857) },
	year         = 2026,
	url          = { https://huggingface.co/datasets/HelmholtzAI-FZJ/mdRNAcat },
	doi          = { 10.57967/hf/11233 },
	publisher    = { Hugging Face }
}

@article{JURECA,
author = {{J\"{u}lich Supercomputing Centre}},
title = {{JURECA: Data Centric and Booster Modules implementing the Modular Supercomputing Architecture at J\"{u}lich Supercomputing Centre}},
journal = {Journal of large-scale research facilities},
number = {A182},
volume = {7},
doi = {10.17815/jlsrf-7-182},
url = {http://dx.doi.org/10.17815/jlsrf-7-182},
year = {2021}
}
\bibliographystyle{iclr2027_conference}

\newpage
\appendix
\section{Appendix}
\subsection{Units used in mdRNAcat} \label{appendix:units}

We note that mdRNAcat intentionally uses different unit conventions across data fields. This choice primarily reflects compatibility with mdCATH  \cite{mirarchi2024mdcath} and with the native output conventions of the simulation software. Coordinates and forces follow the mdCATH convention and are reported in \unit{\angstrom} and \unit{\kilo\calorie\per\mole\per\angstrom}, respectively, whereas replica-exchange quantities retain the corresponding GROMACS conventions, such as \unit{\kilo\joule\per\mole} for energies. The unit of every field is explicitly specified in the dataset schema. Quantities representing the same physical observable can be converted between these conventions by standard unit conversions, which are readily supported by common molecular-simulation libraries.

\subsection{RNA System in mdRNAcat} \label{appendix:pdb_list}

In this section, we list all systems that were considered for the mdRNAcat as a result of the initial Nucleoseeker-query in Table \ref{table:rna-systems}. Additionally the 5 NMR-resolved systems were added. The table contains information on the simulation length of the systems used in mdRNAcat and information on the exclusion reason if not.

\begin{longtable}{l r l l}
\caption{RNA systems considered or used in mdRNAcat.}\\
\label{table:rna-systems}\\
\toprule
PDB ID & Residues & Simulation length & Status \\
\midrule
\endfirsthead
\multicolumn{4}{c}{\tablename\ \thetable\ -- continued}\\
\toprule
PDB ID & Residues & Simulation length & Status \\
\midrule
\endhead
\midrule \multicolumn{4}{r}{\emph{continued on next page}}\\
\endfoot
\bottomrule
\endlastfoot
1duh & 44 & \qty{200}{\nano\second} & Part of mdRNAcat \\
1ehz & 76 &  & Modified residues not parameterized \\
1hr2 & 158 &  & Too large \\
1kh6 & 53 &  & Modified residues not parameterized \\
1kxk & 70 & \qty{300}{\nano\second} & Part of mdRNAcat \\
1nbs & 155 &  & Missing residues in the pdb \\
1u9s & 161 &  & Missing residues in the pdb \\
1vtq & 75 &  & Modified residues not parameterized \\
1xjr & 47 & \qty{300}{\nano\second} & Part of mdRNAcat \\
1z43 & 101 &  & System size out of scope \\
2h0s & 125 &  & Missing residues in the pdb \\
2il9 & 142 &  & Missing residues in the pdb \\
2m8k & 48 & \qty{200}{\nano\second} & Part of mdRNAcat; Added from NMR structure \\
2mhi & 53 & \qty{200}{\nano\second} & Part of mdRNAcat; Added from NMR structure \\
2mqt & 68 & \qty{200}{\nano\second} & Part of mdRNAcat; Added from NMR structure \\
2qus & 69 & \qty{200}{\nano\second} & Part of mdRNAcat \\
2qwy & 52 & \qty{300}{\nano\second} & Part of mdRNAcat \\
3cw5 & 77 &  & Modified residues not parameterized \\
3d2v & 77 & \qty{200}{\nano\second} & Part of mdRNAcat \\
3e5c & 53 & \qty{200}{\nano\second} & Part of mdRNAcat \\
3egz & 65 &  & In scope; not prioritized \\
3npq & 54 &  & Missing residues in the pdb \\
3oxj & 88 &  & System size out of scope \\
3pdr & 161 &  & System size out of scope \\
3q3z & 75 & \qty{200}{\nano\second} & Part of mdRNAcat \\
3r4f & 66 & \qty{200}{\nano\second} & Part of mdRNAcat \\
3rg5 & 86 &  & In scope; not prioritized \\
3slq & 68 & \qty{200}{\nano\second} & Part of mdRNAcat \\
3suh & 101 &  & System size out of scope \\
4enc & 52 & \qty{200}{\nano\second} & Part of mdRNAcat \\
4fe5 & 67 & \qty{200}{\nano\second} & Part of mdRNAcat \\
4frg & 84 & \qty{200}{\nano\second} & Part of mdRNAcat \\
4gxy & 172 &  & Missing residues in the pdb \\
4jf2 & 77 & \qty{200}{\nano\second} & Part of mdRNAcat \\
4jrc & 57 &  & In scope; not prioritized \\
4k27 & 55 & \qty{200}{\nano\second} & Part of mdRNAcat \\
4kqy & 119 &  & System size out of scope \\
4lvw & 89 & \qty{200}{\nano\second} & Part of mdRNAcat \\
4oji & 54 &  & Modified residues not parameterized \\
4p5j & 86 &  & System size out of scope \\
4pqv & 68 &  & System size out of scope \\
4qln & 125 &  & Missing residues in the pdb \\
4r4p & 186 &  & System size out of scope \\
4rum & 94 &  & Missing residues in the pdb \\
4tzx & 71 & \qty{200}{\nano\second} & Part of mdRNAcat \\
4wfl & 107 &  & System size out of scope \\
4xwf & 64 & \qty{200}{\nano\second} & Part of mdRNAcat \\
4y1m & 107 &  & System size out of scope \\
4yaz & 84 &  & In scope; not prioritized \\
4zc7 & 92 &  & System size out of scope \\
5btp & 75 &  & Missing residues in the pdb \\
5dqk & 48 &  & System size out of scope \\
5dun & 54 &  & Modified residues not parameterized \\
5fjc & 93 & \qty{200}{\nano\second} & Part of mdRNAcat \\
5lyu & 58 & \qty{200}{\nano\second} & Part of mdRNAcat \\
5m0h & 42 & \qty{200}{\nano\second} & Part of mdRNAcat \\
5ml7 & 96 &  & System size out of scope \\
5npm & 80 &  & System size out of scope \\
5nwq & 38 & \qty{200}{\nano\second} & Part of mdRNAcat \\
5ob3 & 69 & \qty{200}{\nano\second} & Part of mdRNAcat \\
5t83 & 90 &  & Missing residues in the pdb \\
5u3g & 85 &  & System size out of scope \\
5une & 47 & \qty{200}{\nano\second} & Part of mdRNAcat \\
6cb3 & 101 &  & System size out of scope \\
6ck5 & 117 &  & Missing residues in the pdb \\
6cu1 & 81 & \qty{200}{\nano\second} & Part of mdRNAcat \\
6d3p & 45 & \qty{200}{\nano\second} & Part of mdRNAcat \\
6dmc & 102 &  & System size out of scope \\
6dvk & 95 &  & System size out of scope \\
6fz0 & 53 &  & Missing residues in the pdb \\
6gyv & 192 &  & Missing residues in the pdb \\
6jq5 & 82 &  & System size out of scope \\
6lax & 55 & \qty{200}{\nano\second} & Part of mdRNAcat \\
6n2v & 99 &  & System size out of scope \\
6n5o & 126 &  & System size out of scope \\
6ol3 & 112 &  & Missing residues in the pdb \\
6p2h & 69 & \qty{200}{\nano\second} & Part of mdRNAcat \\
6prv & 58 & \qty{200}{\nano\second} & Part of mdRNAcat \\
6qn3 & 50 & \qty{200}{\nano\second} & Part of mdRNAcat \\
6tf0 & 52 & \qty{200}{\nano\second} & Part of mdRNAcat \\
6ugg & 77 & \qty{200}{\nano\second} & Part of mdRNAcat \\
6vmy & 148 &  & System size out of scope \\
6wjr & 112 &  & System size out of scope \\
6xkn & 101 &  & Missing residues in the pdb \\
7d7v & 57 & \qty{200}{\nano\second} & Part of mdRNAcat \\
7d81 & 50 & \qty{200}{\nano\second} & Part of mdRNAcat \\
7elq & 45 & \qty{200}{\nano\second} & Part of mdRNAcat \\
7eog & 48 & \qty{200}{\nano\second} & Part of mdRNAcat \\
7jju & 51 & \qty{200}{\nano\second} & Part of mdRNAcat \\
7jrr & 51 &  & System size out of scope \\
7jrs & 129 &  & System size out of scope \\
7jrt & 134 &  & System size out of scope \\
7k16 & 51 & \qty{200}{\nano\second} & Part of mdRNAcat \\
7kga & 90 &  & System size out of scope \\
7mky & 66 &  & Modified residues not parameterized \\
7mlw & 128 &  & Missing residues in the pdb \\
7oax & 52 & \qty{200}{\nano\second} & Part of mdRNAcat \\
7qr3 & 69 & \qty{200}{\nano\second} & Part of mdRNAcat \\
7uq6 & 86 &  & In scope; not prioritized \\
7v9e & 68 &  & Modified residues not parameterized \\
7vnv & 78 &  & Modified residues not parameterized \\
7wii & 50 & \qty{200}{\nano\second} & Part of mdRNAcat \\
8eyu & 87 & \qty{200}{\nano\second} & Part of mdRNAcat \\
8fcs & 71 & \qty{200}{\nano\second} & Part of mdRNAcat; Added from NMR structure \\
8hb8 & 55 &  & In scope; not prioritized \\
8hzl & 84 &  & System size out of scope \\
8its & 46 &  & System size out of scope \\
8uiw & 128 &  & Missing residues in the pdb \\
8uo6 & 134 &  & System size out of scope \\
9bun & 48 & \qty{200}{\nano\second} & Part of mdRNAcat \\
9obm & 73 & \qty{200}{\nano\second} & Part of mdRNAcat; Added from NMR structure \\
\end{longtable}

\subsection{Sequence and Structural-Motif Coverage}
\label{appendix:coverage_metric}

\paragraph{Reference populations.}
The reference search included RNA polymer entities from X-ray
structures released by 25 September 2026, with sequence lengths
of 40--192 nucleotides and reported resolution
$\leq 3.6$~\AA. RNA entities in complexes were eligible.
Sequences were obtained from the PDB canonical sequence field
where available, with thymine converted to uracil.
Sequences containing remaining non-ACGU symbols were excluded.
Low-complexity filtering excluded sequences satisfying both
a uracil fraction $\geq 0.65$ and a consecutive uracil run
of at least 16 nucleotides.
These restrictions applied to the reference population;
representatives from all 48 mdRNAcat PDB IDs were retained
irrespective of these criteria.

The sequence analysis used 644 distinct sequences:
48 mdRNAcat representatives and 596 evaluation sequences.
Structural descriptors were available for 820 retained rows,
comprising 51 selected rows across all 48 mdRNAcat PDB IDs
and 769 non-selected rows.
Sequence evaluation therefore weights distinct sequences,
whereas structural evaluation weights retained entity rows.
The representations describe deposited sequences and
structures, which may differ from edited simulation inputs.

\paragraph{Locally normalized coverage.}
For each representation, let $\mathcal{R}$ denote the combined
reference and representative population, $\mathcal{M}$ the
mdRNAcat representatives, and $\mathcal{P}$ the non-selected
evaluation population.
For $i\in\mathcal{P}$, define
\begin{equation}
d_i^{\mathrm{rep}}=\min_{m\in\mathcal{M}}d(i,m),
\qquad
s_i=\operatorname{median}_{j\in\mathcal{N}_{10}(i;\mathcal{R})}
d(i,j),
\qquad
r_i=\frac{d_i^{\mathrm{rep}}}{s_i},
\end{equation}
where $\mathcal{N}_{10}(i;\mathcal{R})$ contains the ten nearest
non-self neighbours in the combined population.
Representatives are excluded from evaluation but remain
eligible neighbours.
The implementation excludes nonfinite ratios, including
those arising from zero local scales.
Writing $\mathcal{P}^{*}$ for the remaining evaluation rows,
coverage is
\begin{equation}
C(t)=\frac{1}{|\mathcal{P}^{*}|}
\sum_{i\in\mathcal{P}^{*}}\mathbf{1}[r_i\leq t].
\end{equation}
Coverage thus refers to evaluable rows rather than necessarily
the complete reference population.
Pointwise 95\% confidence intervals are the 2.5th and 97.5th
percentiles from 300 bootstrap resamples of the fixed
evaluation ratios; feature fitting and neighbour distances
are not recomputed during resampling.

\paragraph{Sequence representation.}
Contiguous nucleotide 3-mer counts were transformed using
TF--IDF with smoothed inverse-document-frequency weights
and row-wise $\ell_2$ normalization.
The resulting vectors were projected using truncated SVD,
retaining $\min(80,\max(8,V-1))$ components, where $V$ is
the observed vocabulary size.
For a complete canonical 3-mer vocabulary, this gives
63 components.
Distances were Euclidean in this SVD representation.
TF--IDF and SVD were fitted to the combined population.

\paragraph{Structural-motif representation.}
RNA 3D Hub loop annotations provided 13 descriptors:
hairpin, internal-loop and junction count densities;
three corresponding residue fractions;
the mean and standard deviation of motif lengths for
each class; and maximum junction order.
Pseudoknot, stem and unpaired-residue descriptors were not
included in the retained feature set.

In the current converter, chain length denotes the number
of distinct residues occurring in downloaded loop annotations,
rather than full polymer length.
Residue fractions use this annotated-residue denominator.
For chains mapped to the same entity, motif counts are summed,
annotated lengths and residue fractions are aggregated by
their maxima, and motif-length statistics by their means.
Count densities divide the aggregated counts by the
aggregated annotated length.
Consequently, these descriptors can depend on chain
multiplicity and annotation availability.

Missing descriptor values were replaced by feature medians.
The imputed features were standardized using means and
standard deviations fitted to the combined structural
population, and Euclidean distances were used for coverage.
A sensitivity analysis additionally used PCA-whitened
features retaining at least 95\% of standardized variance.

\paragraph{Descriptor shifts and visualization.}
For descriptor $k$, the plotted shift is
\begin{equation}
\Delta_k=
\frac{\mu_{k,\mathcal{M}}-\mu_{k,\mathcal{P}}}
{\sigma_{k,\mathcal{P}}},
\end{equation}
using available, non-imputed descriptor values and the
population standard deviation of non-selected rows.
Selected entity rows are weighted equally; this is not
a one-vector-per-PDB comparison.

UMAP projections are used only for visualization.
Background surfaces smooth the precomputed distance ratios
in the embedding plane and use separate color scales.
Structural ratios are clipped at the 4th and 92nd percentiles
for display only.
Neither this clipping nor UMAP affects the coverage curves.
The analysis characterizes deposited structural motifs,
not conformational coverage of the REST2 ensembles.

\subsection{Simulation Parameters} \label{appendix:simulation_params}

This section contains the GROMACS parameters used for each step done in the pre-processing and simulations for mdRNAcat. All simulations were prepared and executed with the same protocols apart from simulation length. For the pre-processing, firstly, energy minimization was done with the parameters from Table \ref{table:minimization_params}. Then NVT equilibration with Table \ref{table:nvt_params} and lastly NPT equilibration with Table \ref{table:npt_params} were done. For the subsequent simulations, all parameters can be found in Table \ref{table:production_params}.

\begin{table}[h]
\centering
\caption{Parameters used for energy minimization.}
\label{table:minimization_params}
\begin{tabular}{ll}
\toprule
Parameter & Value \\
\midrule
integrator     & steep \\
nsteps         & 100000 \\
emtol          & 10 \\
emstep         & 0.01 \\
nstlog         & 1000 \\
nstenergy      & 100 \\
energygrps     & System \\
cutoff-scheme  & Verlet \\
nstlist        & 1 \\
rlist          & 1 \\
pbc            & xyz \\
coulombtype    & PME \\
pme\_order      & 4 \\
vdwtype        & Cut-off \\
vdw-modifier   & Potential-shift-Verlet \\
rcoulomb       & 1.2 \\
rvdw           & 1.2 \\
epsilon\_r      & 1 \\
fourierspacing & 0.12 \\
\bottomrule
\end{tabular}
\end{table}

\begin{table}[h]
\centering
\caption{Parameters used in the NVT equilibration.}
\label{table:nvt_params}
\begin{tabular}{ll}
\toprule
Parameter & Value \\
\midrule
define                & -DPOSRES \\
integrator            & md \\
nsteps                & 50000 \\
dt                    & 0.002 \\
nstxout               & 500 \\
nstvout               & 500 \\
nstenergy             & 500 \\
nstlog                & 500 \\
continuation          & no \\
constraint\_algorithm  & lincs \\
constraints           & h-bonds \\
lincs\_iter            & 1 \\
lincs\_order           & 4 \\
cutoff-scheme         & Verlet \\
nstlist               & 10 \\
rcoulomb              & 1.2 \\
rvdw                  & 1.2 \\
DispCorr              & EnerPres \\
coulombtype           & PME \\
pme\_order             & 4 \\
fourierspacing        & 0.12 \\
tcoupl                & Nose-Hoover \\
tc-grps               & RNA Water Ion \\
tau\_t                 & 0.5 0.5 0.5 \\
ref\_t                 & 300 300 300 \\
pcoupl                & no \\
pbc                   & xyz \\
gen\_vel               & yes \\
gen\_temp              & 300 \\
gen\_seed              & -1 \\
\bottomrule
\end{tabular}
\end{table}

\begin{table}[h]
\centering
\caption{Parameters used for NPT equilibration.}
\label{table:npt_params}
\begin{tabular}{ll}
\toprule
Parameter & Value \\
\midrule
define                & -DPOSRES \\
integrator            & md \\
nsteps                & 50000 \\
dt                    & 0.002 \\
nstxout               & 500 \\
nstvout               & 500 \\
nstenergy             & 500 \\
nstlog                & 500 \\
continuation          & yes \\
constraint\_algorithm  & lincs \\
constraints           & h-bonds \\
lincs\_iter            & 1 \\
lincs\_order           & 4 \\
cutoff-scheme         & Verlet \\
nstlist               & 10 \\
rcoulomb              & 1.2 \\
rvdw                  & 1.2 \\
DispCorr              & EnerPres \\
coulombtype           & PME \\
pme\_order             & 4 \\
fourierspacing        & 0.12 \\
tcoupl                & Nose-Hoover \\
tc-grps               & RNA Water Ion \\
tau\_t                 & 0.5 0.5 0.5 \\
ref\_t                 & 300 300 300 \\
pcoupl                & Parrinello-Rahman \\
pcoupltype            & semiisotropic \\
tau\_p                 & 2.0 \\
ref\_p                 & 1.0 1.0 \\
compressibility       & 4.5e-5 4.5e-5 \\
refcoord\_scaling      & com \\
pbc                   & xyz \\
gen\_vel               & no \\
\bottomrule
\end{tabular}
\end{table}

\begin{table}[h]
\centering
\caption{Parameters used for production simulation runs.}
\label{table:production_params}
\begin{tabular}{ll}
\toprule
Parameter & Value \\
\midrule
integrator            & md \\
nsteps                & 100000000 \\
dt                    & 0.002 \\
nstxout               & 250 \\
nstvout               & 250 \\
nstfout               & 250 \\
nstenergy             & 250 \\
nstlog                & 5000 \\
nstcalcenergy         & 250 \\
energygrps            & RNA Water Ion \\
continuation          & yes \\
constraint\_algorithm  & lincs \\
constraints           & h-bonds \\
lincs\_iter            & 1 \\
lincs\_order           & 4 \\
cutoff-scheme         & Verlet \\
nstlist               & 10 \\
coulombtype           & PME \\
pme\_order             & 4 \\
rlist                 & 1.2 \\
rcoulomb              & 1.2 \\
rvdw                  & 1.2 \\
fourierspacing        & 0.12 \\
tcoupl                & Nose-Hoover \\
tc-grps               & RNA Water Ion \\
tau\_t                 & 0.5 0.5 0.5 \\
ref\_t                 & 300 300 300 \\
pcoupl                & Parrinello-Rahman \\
pcoupltype            & isotropic \\
tau\_p                 & 2.0 \\
ref\_p                 & 1.0 1.0 \\
compressibility       & 4.5e-5 4.5e-5 \\
pbc                   & xyz \\
DispCorr              & EnerPres \\
gen\_vel               & no \\
nstcomm               & 100 \\
comm-grps             & RNA Water Ion \\
\bottomrule
\end{tabular}
\end{table}

\begin{table}[h]
\centering
\caption{REST2 effective-temperature ladder and $\lambda$ scaling factors.}
\label{tab:rest2_lambda}
\begin{tabular}{c c c c}
\toprule
Replica & $T_i$ [\unit{\kelvin}] & $\lambda$ & $\sqrt{\lambda}$ \\
\midrule
1 & 300.000000 & 1.000000 & 1.000000 \\
2 & 304.980000 & 0.983700 & 0.991817 \\
3 & 310.050000 & 0.967600 & 0.983667 \\
4 & 315.200000 & 0.951800 & 0.975602 \\
5 & 320.440000 & 0.936200 & 0.967574 \\
6 & 325.760000 & 0.920900 & 0.959635 \\
7 & 331.180000 & 0.905900 & 0.951788 \\
8 & 336.680000 & 0.891100 & 0.943981 \\
9 & 342.270000 & 0.876500 & 0.936216 \\
10 & 347.960000 & 0.862200 & 0.928547 \\
11 & 353.740000 & 0.848100 & 0.920923 \\
12 & 359.620000 & 0.834200 & 0.913345 \\
13 & 365.590000 & 0.820600 & 0.905870 \\
14 & 371.670000 & 0.807200 & 0.898443 \\
15 & 377.840000 & 0.794000 & 0.891067 \\
16 & 384.120000 & 0.781000 & 0.883742 \\
17 & 390.500000 & 0.768200 & 0.876470 \\
18 & 396.990000 & 0.755700 & 0.869310 \\
19 & 403.590000 & 0.743300 & 0.862148 \\
20 & 410.290000 & 0.731200 & 0.855102 \\
21 & 417.110000 & 0.719200 & 0.848057 \\
22 & 424.040000 & 0.707500 & 0.841130 \\
23 & 431.080000 & 0.695900 & 0.834206 \\
24 & 438.250000 & 0.684500 & 0.827345 \\
25 & 445.530000 & 0.673400 & 0.820610 \\
26 & 452.930000 & 0.662400 & 0.813880 \\
27 & 460.460000 & 0.651500 & 0.807155 \\
28 & 468.110000 & 0.640900 & 0.800562 \\
29 & 475.880000 & 0.630400 & 0.793977 \\
30 & 483.790000 & 0.620100 & 0.787464 \\
31 & 491.830000 & 0.610000 & 0.781025 \\
32 & 500.000000 & 0.600000 & 0.774597 \\
\bottomrule
\end{tabular}
\end{table}

\FloatBarrier 
\subsection{Energy Overlap Plots} \label{appendix:overlap_plots}

\begin{figure}[h]
    \centering
    \includegraphics[width=0.9\textwidth]{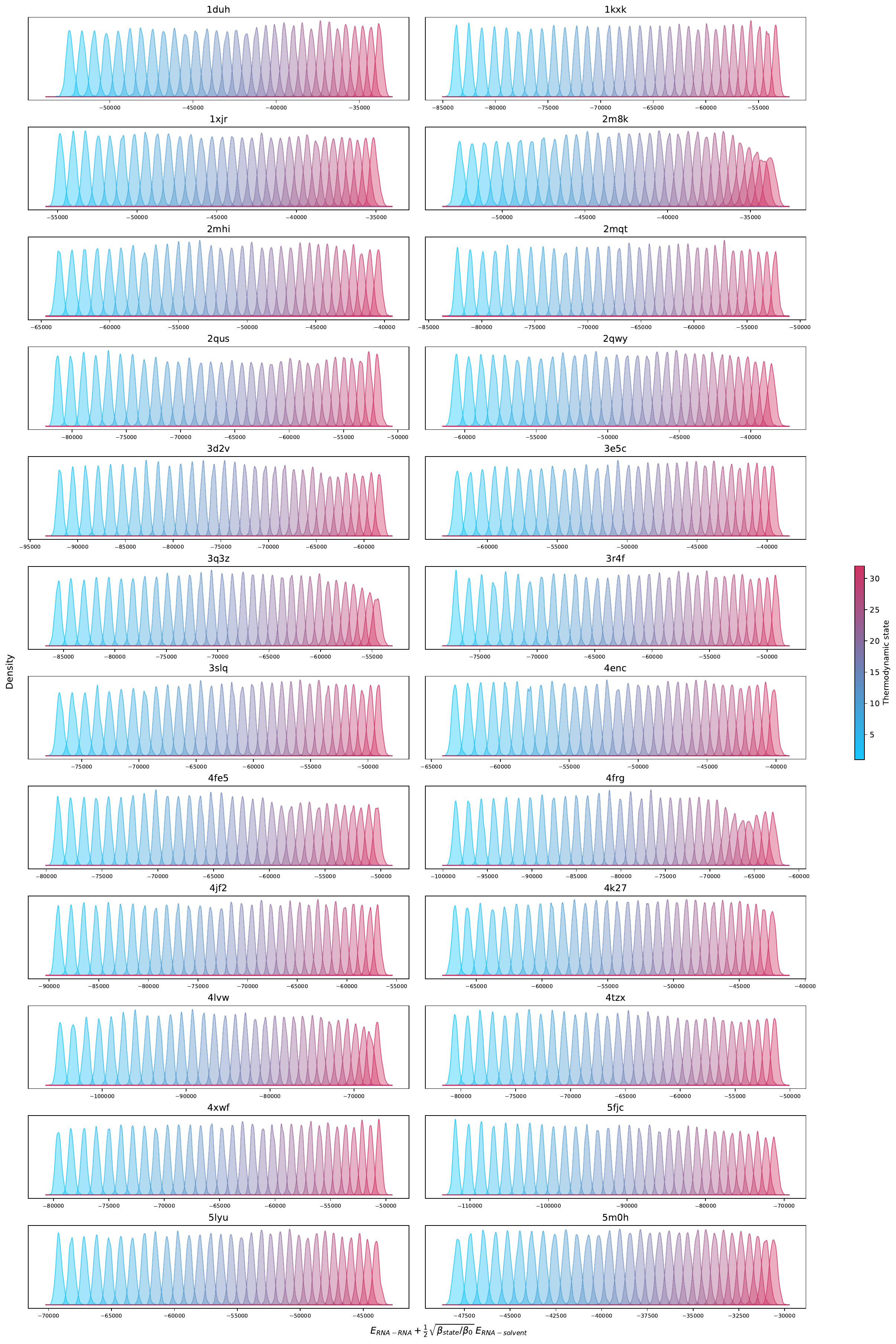}
    \caption{Density of the thermodynamic states against the energy of the first 24 systems in mdRNAcat.}
    \label{plot:repex_overlap_1}
\end{figure}

\begin{figure}[h]
    \centering
    \includegraphics[width=0.9\textwidth]{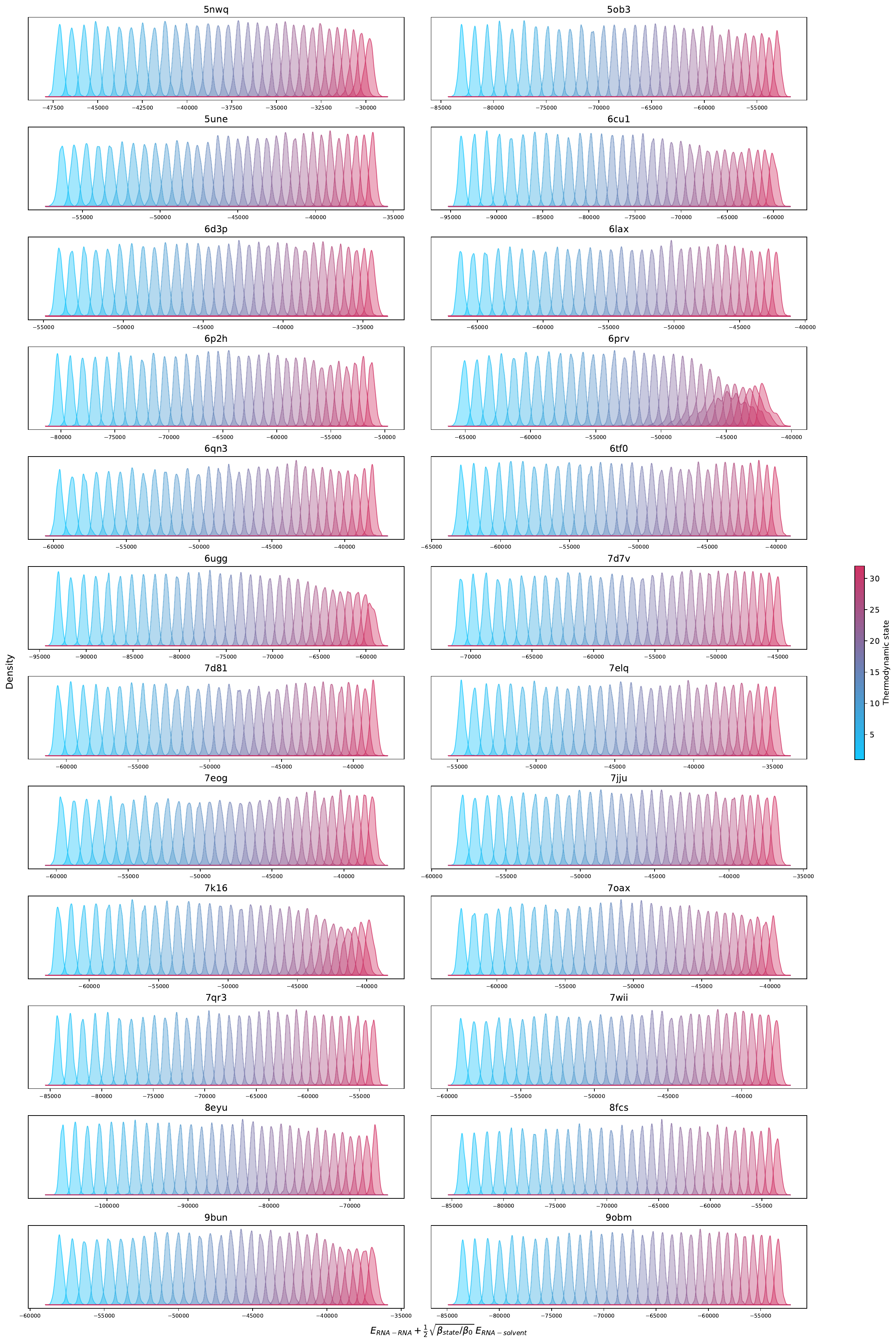}
    \caption{Density of the thermodynamic states against the energy of the second 24 systems in mdRNAcat.}
    \label{plot:repex_overlap_2}
\end{figure}

This section contains the energy overlap plot for all 48 simulations that were done for mdRNAcat in Figures \ref{plot:repex_overlap_1} and \ref{plot:repex_overlap_2}.


\end{document}